\documentclass[twocolumn]{aastex63}

\usepackage{amsmath}	
\usepackage{enumitem}
\setlist[enumerate]{leftmargin=\parindent}
\setlist[itemize]{leftmargin=\parindent}
\setlist{nosep}
\usepackage{lipsum}
\usepackage[figuresright]{rotating}
\usepackage{booktabs}
\usepackage{changepage}

\newcommand{\Kepler}{\emph{Kepler}}
\newcommand{\TESS}{\emph{TESS}}

\newcommand{\Msun}{\mbox{$M_{\odot}$}}
\newcommand{\Rsun}{\mbox{$R_{\odot}$}}

\newcommand\revision[1]{\textcolor{black}{#1}}

\makeatletter
\newcommand\notsotiny{\@setfontsize\notsotiny\@vipt\@viipt}
\makeatother

\begin{document}

\vspace*{-1.3in}
\title[Complex Modulation of Rapidly Rotating Young M Dwarfs]{Complex Modulation of Rapidly Rotating Young M Dwarfs: Adding Pieces to the Puzzle}
\shorttitle{Complex Rotators II}

\email{maximilian.guenther@esa.int}

\author[0000-0002-3164-9086]{Maximilian N. G{\"u}nther}
\thanks{Juan Carlos Torres Fellow, \revision{ESA Research Fellow}}
\affiliation{\scriptsize Department of Physics, and Kavli Institute for Astrophysics and Space Research, MIT, 77 Mass. Ave, Cambridge, MA 02139, USA}
\affiliation{\scriptsize \revision{European Space Agency (ESA), European Space Research and Technology Centre (ESTEC), Keplerlaan 1, 2201 AZ Noordwijk, Netherlands}}


\author[0000-0001-6298-412X]{David A. Berardo}
\affiliation{\scriptsize Department of Physics, and Kavli Institute for Astrophysics and Space Research, MIT, 77 Mass. Ave, Cambridge, MA 02139, USA}

\author[0000-0001-6298-412X]{Elsa Ducrot}
\affiliation{\scriptsize Astrobiology Research Unit, Universit{\'e} de Li{\`e}ge, 19C Allée du 6 Ao\t{u}t, 4000 Li{\`e}ge, Belgium}

\author[0000-0001-8504-5862]{Catriona A. Murray}
\affiliation{\scriptsize Cavendish Laboratory, JJ Thomson Avenue, Cambridge CB3 0HE, UK}

\author[0000-0002-3481-9052]{Keivan G. Stassun}
\affiliation{\scriptsize Department of Physics and Astronomy, Vanderbilt University, Nashville, TN 37235, USA}

\author[0000-0003-3669-7201]{Katalin Olah}
\affiliation{\scriptsize Konkoly Observatory, Research Centre for Astronomy and
Earth Sciences, Hungary}

\author[0000-0002-0514-5538]{L.G. Bouma}
\affiliation{\scriptsize Department of Astrophysical Sciences, Princeton University, 4 Ivy Lane, Princeton, NJ 08544, USA}

\author{Saul Rappaport}
\affiliation{\scriptsize Department of Physics, and Kavli Institute for Astrophysics and Space Research, MIT, 77 Mass. Ave, Cambridge, MA 02139, USA}

\author{Joshua N. Winn}
\affiliation{\scriptsize Department of Astrophysical Sciences, Princeton University, 4 Ivy Lane, Princeton, NJ 08544, USA}

\author[0000-0002-9464-8101]{Adina D. Feinstein}
\thanks{NSF Graduate Research Fellow}
\affiliation{\scriptsize Department of Astronomy and Astrophysics, University of
Chicago, 5640 S. Ellis Ave, Chicago, IL 60637, USA}

\author{Elisabeth \revision{C.} Matthews}
\affiliation{\scriptsize Department of Physics, and Kavli Institute for Astrophysics and Space Research, MIT, 77 Mass. Ave, Cambridge, MA 02139, USA}

\author{Daniel Sebastian}
\affiliation{\scriptsize Astrobiology Research Unit, Universit{\'e} de Li{\`e}ge, 19C Allée du 6 Ao\t{u}t, 4000 Li{\`e}ge, Belgium}

\author[0000-0002-3627-1676]{Benjamin V. Rackham}
\thanks{51 Pegasi b Fellow}
\affiliation{\scriptsize Department of Physics, and Kavli Institute for Astrophysics and Space Research, MIT, 77 Mass. Ave, Cambridge, MA 02139, USA}
\affiliation{\scriptsize Department of Earth, Atmospheric and Planetary Sciences, MIT, 77 Mass. Ave, Cambridge, MA 02139, USA}

\author[0000-0002-3658-2175]{B{\'a}lint Seli}
\affiliation{\scriptsize Konkoly Observatory, Research Centre for Astronomy and
Earth Sciences, Hungary}

\author[0000-0002-5510-8751]{Amaury H. M. J. Triaud}
\affiliation{\scriptsize School of Physics \& Astronomy, University of Birmingham, Edgbaston, Birmimgham B15 2TT, UK}

\author{Edward Gillen}
\thanks{Winton Fellow}
\affiliation{\scriptsize Astronomy Unit, Queen Mary University of London, Mile End Road, London E1 4NS, UK}
\affiliation{\scriptsize Cavendish Laboratory, JJ Thomson Avenue, Cambridge CB3 0HE, UK}

\author[0000-0001-8172-0453]{Alan M. Levine}
\affiliation{\scriptsize Department of Physics, and Kavli Institute for Astrophysics and Space Research, MIT, 77 Mass. Ave, Cambridge, MA 02139, USA}

\author[0000-0002-9355-5165]{Brice-Olivier Demory}
\affiliation{\scriptsize University of Bern, Center for Space and Habitability, Bern, Switzerland}

\author{Micha\"el Gillon}
\affiliation{\scriptsize Astrobiology Research Unit, Universit{\'e} de Li{\`e}ge, 19C Allée du 6 Ao\t{u}t, 4000 Li{\`e}ge, Belgium}

\author{Didier Queloz}
\affiliation{\scriptsize Cavendish Laboratory, JJ Thomson Avenue, Cambridge CB3 0HE, UK}
\affiliation{\scriptsize Observatoire astronomique de l’Université de Genève, 51 chemin de Pégase, 1290 Sauverny, Switzerland}


\author{George R. Ricker}
\affiliation{\scriptsize Department of Physics, and Kavli Institute for Astrophysics and Space Research, MIT, 77 Mass. Ave, Cambridge, MA 02139, USA}

\author{Roland K. Vanderspek}
\affiliation{\scriptsize Department of Physics, and Kavli Institute for Astrophysics and Space Research, MIT, 77 Mass. Ave, Cambridge, MA 02139, USA}

\author[0000-0002-6892-6948]{Sara Seager}
\affiliation{\scriptsize Department of Physics, and Kavli Institute for Astrophysics and Space Research, MIT, 77 Mass. Ave, Cambridge, MA 02139, USA}
\affiliation{\scriptsize Department of Earth, Atmospheric and Planetary Sciences, MIT, 77 Mass. Ave, Cambridge, MA 02139, USA}
\affiliation{\scriptsize Department of Aeronautics and Astronautics, MIT, 77 Mass. Ave, Cambridge, MA 02139, USA}

\author{David W. Latham}
\affiliation{\scriptsize Center for Astrophysics | Harvard \& Smithsonian, 60 Garden Street, Cambridge, MA 02138}

\author{Jon M. Jenkins}
\affiliation{\scriptsize NASA Ames Research Center, Moffett Field, CA, 94035, USA}

\author[0000-0002-9314-960X]{C.~E.~Brasseur} 
\affiliation{\scriptsize Mikulski Archive for Space Telescopes}




\author[0000-0001-8020-7121]{Knicole D. Col\'{o}n} 
\affiliation{\scriptsize NASA Goddard Space Flight Center, Exoplanets and Stellar Astrophysics Laboratory (Code 667), Greenbelt, MD 20771, USA}         

\author[0000-0002-6939-9211]{Tansu Daylan} 
\thanks{Kavli Fellow}
\affiliation{\scriptsize Department of Physics, and Kavli Institute for Astrophysics and Space Research, MIT, 77 Mass. Ave, Cambridge, MA 02139, USA}

\author[0000-0001-6108-4808]{Laetitia Delrez} 
\affiliation{\scriptsize Astrobiology Research Unit, Universit{\'e} de Li{\`e}ge, 19C Allée du 6 Ao\t{u}t, 4000 Li{\`e}ge, Belgium}
\affiliation{\scriptsize Space Sciences, Technologies and Astrophysics Research (STAR) Institute, Universit{\'e} de Li{\`e}ge, 19C All{\'e}e du 6 Ao\t{u}t, 4000 Li{\`e}ge, Belgium}
\affiliation{\scriptsize Observatoire astronomique de l’Université de Genève, 51 chemin de Pégase, 1290 Sauverny, Switzerland}

\author[0000-0002-9113-7162]{\revision{Michael Fausnaugh}} 
\affiliation{\scriptsize Department of Physics, and Kavli Institute for Astrophysics and Space Research, MIT, 77 Mass. Ave, Cambridge, MA 02139, USA}


\author{Lionel J. Garcia} 
\affiliation{\scriptsize Astrobiology Research Unit, Universit{\'e} de Li{\`e}ge, 19C Allée du 6 Ao\t{u}t, 4000 Li{\`e}ge, Belgium}




\author[0000-0002-7778-3117]{Rahul Jayaraman}
\affiliation{\scriptsize Department of Physics, and Kavli Institute for Astrophysics and Space Research, MIT, 77 Mass. Ave, Cambridge, MA 02139, USA}

\author{Emmanuel Jehin} 
\affiliation{\scriptsize Space Sciences, Technologies and Astrophysics Research (STAR) Institute, Universit{\'e} de Li{\`e}ge, 19C All{\'e}e du 6 Ao\t{u}t, 4000 Li{\`e}ge, Belgium}

\author[0000-0002-2607-138X]{Martti H. Kristiansen} 
\affiliation{\scriptsize Brorfelde Observatory, Observator Gyldenkernes Vej 7, DK-4340 T\o{}ll\o{}se, Denmark}
\affiliation{\scriptsize DTU Space, National Space Institute, Technical University of Denmark, Elektrovej 327, DK-2800 Lyngby, Denmark}

\author[0000-0002-8804-0212]{J. M. Diederik Kruijssen} 
\affiliation{\scriptsize Astronomisches Rechen-Institut, Zentrum f\"ur Astronomie der Universit\"at Heidelberg, M\"onchhofstra\ss e 12-14, D-69120 Heidelberg, Germany}



\author[0000-0002-5220-609X]{Peter Pihlmann Pedersen} 
\affiliation{\scriptsize Cavendish Laboratory, JJ Thomson Avenue, Cambridge CB3 0HE, UK}

\author{Francisco J. Pozuelos} 
\affiliation{\scriptsize Space Sciences, Technologies and Astrophysics Research (STAR) Institute, Universit{\'e} de Li{\`e}ge, 19C All{\'e}e du 6 Ao\t{u}t, 4000 Li{\`e}ge, Belgium}
\affiliation{\scriptsize Astrobiology Research Unit, Universit{\'e} de Li{\`e}ge, 19C Allée du 6 Ao\t{u}t, 4000 Li{\`e}ge, Belgium}


\author[0000-0001-8812-0565]{Joseph E. Rodriguez} 
\affiliation{\scriptsize Center for Astrophysics | Harvard \& Smithsonian, 60 Garden Street, Cambridge, MA 02138}



\author[0000-0002-5402-9613]{Bill Wohler} 
\affiliation{\scriptsize SETI Institute/NASA Ames Research Center}

\author[0000-0002-4142-1800]{Zhuchang Zhan}
\affiliation{\scriptsize Department of Earth, Atmospheric and Planetary Sciences, MIT, 77 Mass. Ave, Cambridge, MA 02139, USA}

\begin{abstract}

\revision{\small 
New sets of young M dwarfs with complex, sharp-peaked, and strictly periodic photometric modulations have recently been discovered with Kepler/K2 (scallop shells) and TESS (complex rotators). All are part of star-forming associations, are distinct from other variable stars, and likely belong to a unified class. Suggested hypotheses include star spots, accreting dust disks, co-rotating clouds of material, magnetically constrained material, spots and misaligned disks, and pulsations. Here, we provide a comprehensive overview and add new observational constraints with TESS and SPECULOOS Southern Observatory (SSO) photometry. We scrutinize all hypotheses from three new angles: (1) we investigate each scenario's occurrence rates via young star catalogs; (2) we study the features’ longevity using over one year of combined data; and (3) we probe the expected color dependency with multi-color photometry. In this process, we also revisit the stellar parameters accounting for activity effects, study stellar flares as activity indicators over year-long time scales, and develop toy models to simulate typical morphologies. We rule out most hypotheses, and only (i) co-rotating material clouds and (ii) spots and misaligned disks remain feasible - with caveats. For (i), co-rotating dust might not be stable enough, while co-rotating gas alone likely cannot cause percentage-scale features; and (ii) would require misaligned disks around most young M dwarfs. We thus suggest a unified hypothesis, a superposition of large-amplitude spot modulations and sharp transits of co-rotating gas clouds. While the complex rotators' mystery remains, these new observations add valuable pieces to the puzzle going forward.
}
\end{abstract}


\section{Introduction}
\label{s:Introduction}

\subsection{Morphologies of young M dwarfs}

\revision{Young M dwarf stars (here 20--150\,Myr) are often fast rotators, with rotational periods ranging from hours to one or two days.
Their rotation is one of the drivers of their magnetic dynamos and thus stellar activity \citep[e.g.,][]{Moffatt1978, Parker1979, Browning2008}.}
This can be observed in terms of activity indicators, such as hydrogen and calcium H\,\&\,K emission lines, frequent and strong flaring activity, and significant star spot coverage  \citep[e.g.,][]{Benz2010, West2015, Newton2017, Wright2018, Guenther2020a}.

\begin{figure*}[!htbp]
    \centering
    \includegraphics[width=0.85\textwidth]{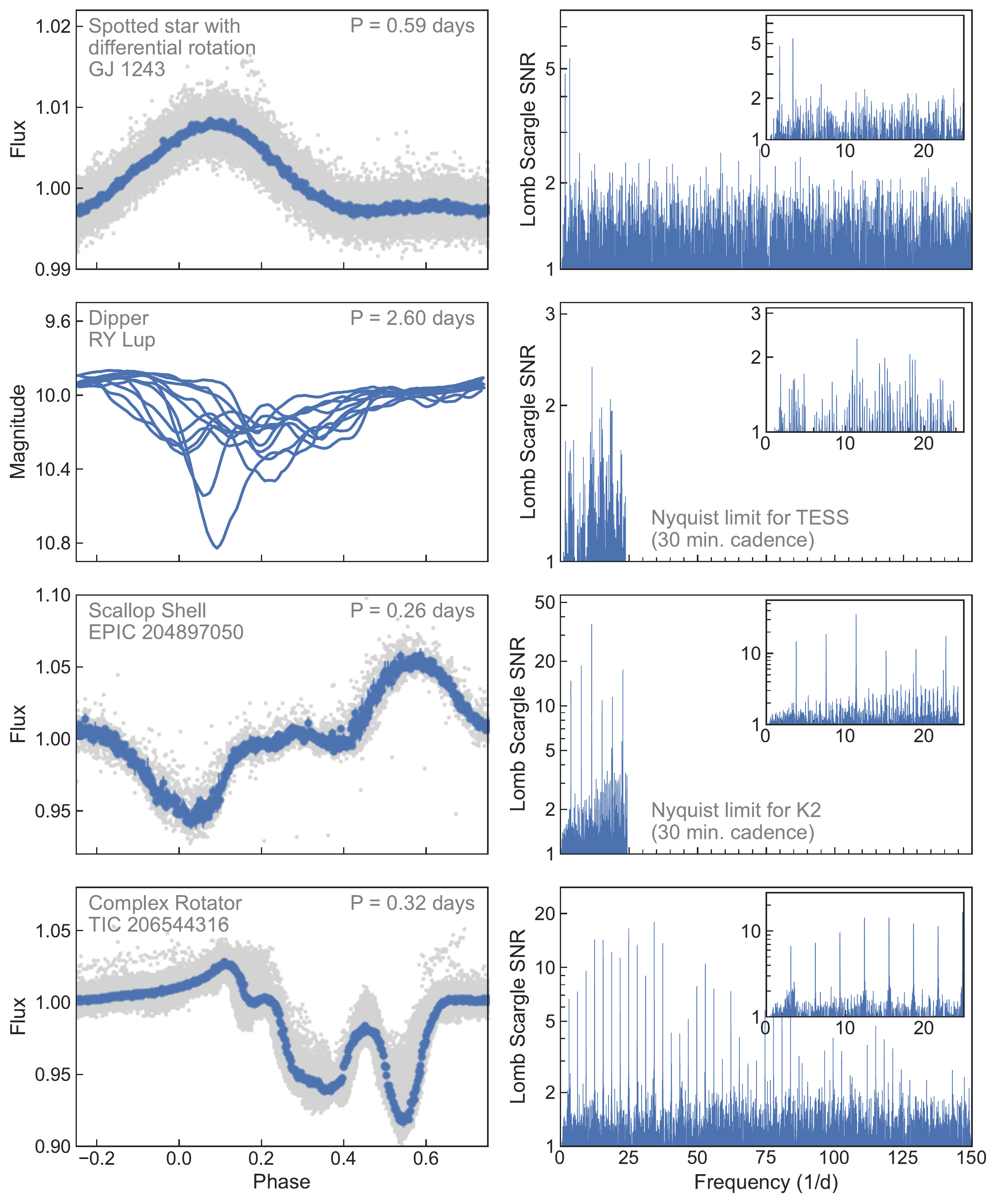}
    \caption{A collage of different morphology classes of young M dwarfs with complex photometric variability. 
    \textbf{First row:} a typical \textit{complex rotator} (TIC~206544316) with a period of 0.32~days. Its phase-folded light curve highlights the complex modulation, and its frequency spectrum shows dozens of harmonics down to time scales of minutes.
    \textbf{Second row:} a typical \textit{scallop shell} (EPIC~204897050) with a period of 0.26~days. It is apparent that these targets share the same morphology class as the \textit{complex rotators}, even though the data are constrained by the Nyquist limit for K2 (30 min. cadence).
    \textbf{Third row:} a typical spotted M dwarf with a period of 0.58~days \citep[GJ~1243; see e.g.][]{Davenport2015, Davenport2020, allesfitter-paper}, showing a simple modulation and only two peaks in the frequency spectrum (two peaks rather than one due to differential rotation). 
    \textbf{Fourth row:} a typical \textit{dipper} star observed with TESS \citep[RY Lup;][]{Bredall2020}, usually showing no strict periodicity, no presence of multiple clear harmonics, and a wider range of periods (multiple days to weeks).
    Not shown are the more standard variability classes, such as single spot modulation and binary stars.
    }
    \label{fig:morphology_classes}
\end{figure*}

In photometric observations, young M dwarfs with spots often show smooth, semi-sinusoidal rotational modulation with amplitudes of a few percent.
Their patterns are rather `simple', manifesting only a few peaks in a Fourier transform, even in the presence of multiple spots and differential rotation (Fig.~\ref{fig:morphology_classes} \revision{first} row).
Thus, even the most extreme rotational modulations discovered so far can be described by just a handful of spots \citep[e.g.,][]{Rappaport2014, Strassmeier2017}.


One of the first phenomena clearly standing out from this norm were \textit{dipper} and \textit{burster} stars. These show abrupt dips or bursts of light in a quasi-periodic or stochastic manner \citep[][Fig.~\ref{fig:morphology_classes} \revision{second} row]{Alencar2010, Morales-Calderon2011, Cody2014, Ansdell2016}, and were grouped by their photometric morphology into seven distinct classes\footnote{
\textit{periodic dippers}, 
\textit{aperiodic dippers},
\textit{stochastic variables},
\textit{periodic variables} (likely spots),
\textit{quasi-periodic variables},
\textit{bursters}, and
\textit{long-timescale variables}.}. 

Shortly after, \citet{Stauffer2017, Stauffer2018} discovered yet again three new morphology classes in Kepler/K2 data\footnote{
\textit{scallop shell}, 
\textit{persistent flux-dip}, and 
\textit{narrow flux-dip} variables}. 
As these three share common features, we refer to them collectively as \textit{scallop shells} throughout this paper (Fig.~\ref{fig:morphology_classes} \revision{third} row).
These \textit{scallop shells} differ from the \textit{dippers/bursters} in two substantial ways: (1) the objects discussed by \citet{Stauffer2018} are strictly periodic; and (2) they rotate much more rapidly, typically on timescales of $\lesssim$2 days, compared to the timescales of multiple days to weeks for the \textit{dippers/bursters}.



\revision{Most recently, \citet{Zhan2019} discovered ten very similar objects in TESS Sectors 1\,\&\,2,, dubbed \textit{complex rotators} (Fig.~\ref{fig:morphology_classes} \revision{fourth} row; see also Fig.~\ref{fig:TESS_lightcurves_collage} for a collage of all light curves per TESS orbit).
All \textit{complex rotators} and \textit{scallop shells} show rapid rotation, strict periodicity, and dozens of harmonics in their frequency spectra indicating their sharp light curve features. We thus argue that they are likely the same class of objects, and any differences are only due to Kepler's and TESS' observing cadences (30\,min vs. 2\,min).
This makes them strictly distinct from `normal' spotted stars (even those with differential rotation), which only show one or two peaks in their frequency spectra, and from \textit{dippers}, which are far less periodic and morphological stable.}

\subsection{Hypotheses for complex modulations and their limitations}

\begin{figure}[!htbp]
    \centering
    \includegraphics[width=\columnwidth]{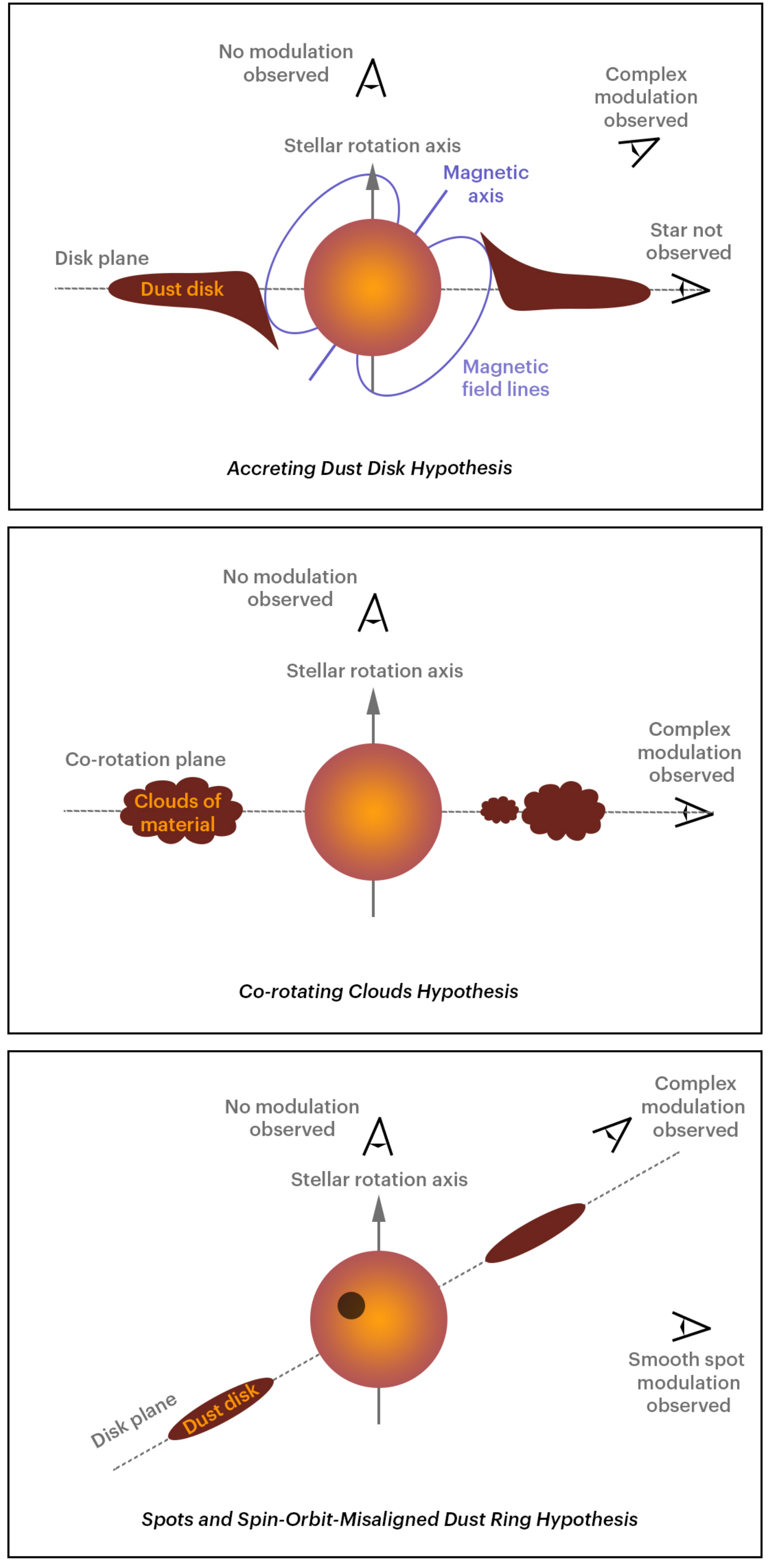}
    \caption{
    \small
    Three hypotheses to explain complex morphologies of young M dwarfs.
    \textbf{First panel:} \textit{Accreting dust disk} hypothesis \citep{Bodman2016}\revision{, where} an accreting dust disk seen from different observing angles could explain \textit{dipper} and \textit{burster} stars.
    \textbf{Second panel:} \textit{Co-rotating clouds} hypothesis \citep{Stauffer2017,Stauffer2018}\revision{, where} a patchy torus of gas clouds at the Keplerian co-rotation orbit periodically blocks out stellar light and might cause \textit{scallop shell} modulation.
    \textbf{Third panel:} \textit{Spots and misaligned disk} hypothesis \citep{Zhan2019}\revision{, where spot occultations might explain the \textit{complex rotator's} modulation}.
    }
    \label{fig:hypotheses}
\end{figure}

\revision{All} \textit{dipper} and \textit{burster} classes were linked to the presence of dusty disks and a viewing-angle dependency, suggested by observations of strong infrared excess \citep[\textit{accreting dust disks};][Fig.~\ref{fig:hypotheses} first panel]{Bodman2016}.

The \textit{scallop shells} were \revision{first} suggested to arise from a patchy torus of material clouds at the Keplerian co-rotation radius periodically transiting the star \citep[\textit{co-rotating clouds};][Fig.~\ref{fig:hypotheses} second panel]{Stauffer2017, Stauffer2018}\footnote{\citet{Yu2015} and \citet{Bouma2020} suggested a similar explanation for the T Tauri star PTFO 8-8695b.}. \revision{Such material might be warm coronal gas, dust, or a mixture of both.}

\revision{When studying the \textit{complex rotators}, however, \citet{Zhan2019} 
suggested a new idea:}
spotted, rapidly rotating stars that host an inner dust disk at a few stellar radii, and show a spin-orbit misalignment between their rotation axis and the dust ring (\textit{spots and misaligned disk}; Fig.~\ref{fig:hypotheses} third panel). \revision{Spots might then pass behind the dust disk and get (partially) occulted, leading to sudden increases in photometric brightness.}


\revision{In particular,} \citet{Zhan2019} presented the following counterarguments to the previous hypotheses:
\begin{itemize}
    \item \textit{Spots only}: 
        even the superposition of \revision{multiple} cold and hot stellar spots leads to smooth variations and can only explain \revision{1--2} peaks in the frequency spectrum \citep[also][]{Kovari1997,Stauffer2017}.
    \item \textit{Accreting dust disk}: 
        \revision{(i) \textit{complex rotators}' stable periodicity is in stark contrast to the semi-periodic and stochastic nature of morphologies caused by accreting disks; 
        (ii) the absence of significant infrared excess in their spectral energy distributions (SEDs) contradicts the presence of accreting disks;
        (iii) their roation periods are much shorter than those of \textit{dippers} \citep[also][]{Stauffer2017}.}
    \item \textit{Co-rotating clouds}: 
        (i) if the material is gas, it is challenging to explain the large amplitudes of the modulation;
        (ii) the material cannot be dust, as it cannot be stably confined at the required distances of several stellar radii because the magnetic field at those large distances from the surface would be too weak;
        (iii) any material (be it gas or dust) trapped in the magnetic field closer to the stellar surface could not reproduce the observations of sharp features with amplitudes of several percent.
\end{itemize}

\vspace{-6pt}
\subsection{This paper}

\revision{This paper focuses on the ten targets discovered by \citet{Zhan2019} to further scrutinize the plausibility of several hypotheses by three means:
\begin{enumerate}[noitemsep]
    \item investigating their occurrence rates,
    \item studying the morphologies' stability and longevity over one (non-continuous) year, and
    \item probing the feature's chromaticity.
\end{enumerate}
}

\revision{We give an overview of all observations in Section \ref{s:Observations}, and} revise the stellar parameters in Section~\ref{s:Revising the stellar parameters}.
Then, we study stellar flares and other brightenings in Section~\ref{s:Flares and sudden brightenings}.
Next, we \revision{scrutinize all} hypotheses with respect to occurrence rates, stability and longevity, and color dependency in Section~\ref{s:Testing the hypotheses}.
Finally, we discuss our findings and present our conclusion in Sections \ref{s:Discussion} and \ref{s:Conclusion}.

\section{Observations}
\label{s:Observations}

\subsection{TESS photometry}
\label{ss:TESS Photometry}

\revision{The ten \textit{complex rotators} were discovered in TESS short-cadence data from Sector 1 (2018-07-25 to 2018-08-22) and Sector 2 (2018-08-22 to 2018-09-10) and observed as part of the cool dwarf catalog \citep{Stassun2018, Muirhead2018}.}
Here, we also add new data taken over the full first year of operations (Tab.~\ref{tab:updated_parameters_of_the_complex_rotators}).
Light curves were prepared with the Science Processing Operations Center (SPOC) pipeline \citep{Jenkins2016}, a descendant of the \Kepler{} mission pipeline \citep{Jenkins2002, Jenkins2010, Jenkins2017, Stumpe2014, Smith2012}. \revision{We use the pre-search data conditioned simple aperture (PDC-SAP) light curves, which are detrended for instrumental systematics.}

\subsection{SPECULOOS Southern Observatory photometry}
\label{ss:SPECULOOS Southern Observatory Photometry}

The SPECULOOS Southern Observatory \citep[SSO;][]{Gillon2018, Burdanov2018, Delrez2018} is located at ESO’s Paranal Observatory in Chile and is part of the SPECULOOS network. The facility consists of four robotic 1-m telescopes (Callisto, Europa, Ganymede, and Io), each equipped with a near-infrared-sensitive CCD camera with a resolution of 0.35 arcsec per pixel.
We observed four targets, TIC~201789285, TIC~206544316, TIC~332517282, and TIC~425933644, each simultaneously in at least two wavelength bands (g', r', i', and z' band filters) for an entire observing night \revision{(Table~\ref{tab:sso_observations})}. 
\revision{We extracted light curves using the SSO pipeline \citep{Murray2020}, which uses the \texttt{casutools} software \citep{Irwin2004} for automated differential photometry and detrends for telluric water vapor}.

\subsection{ANU spectroscopy}
\label{ss:ANU Spectroscopy}

We also reuse the spectroscopic observations taken by \citet{Zhan2019}. The low-resolution spectra covered four of the systems (TIC 177309964, TIC 206544316, TIC 234295610, and TIC 425933644) using the Wide Field Spectrograph \citep[WiFeS;][]{Dopita2007} on the Australian National University (ANU) 2.3\,m telescope at Siding Spring Observatory, Australia, on January 18 and 19, 2019. The observations cover the 5200--7000\,\AA{} band with a resolving power of $R=7000$ and were reduced following \citet{Bayliss2013}. All spectra reveal strong H$\alpha$ emission features with equivalent widths of 4--7\,\AA{}, typical of rapidly rotating young M stars (see Section~\ref{s:Revising the stellar parameters}). \revision{No signs of a binary nature of these four objects were found.}

\section{Revising the stellar parameters}
\label{s:Revising the stellar parameters}
\revision{
\vspace*{-11pt}
\subsection{Revisiting the binary TIC~289840928 and TIC~289840926}
\label{ss:Revisiting the binary TIC289840928 and TIC289840926}
\cite{Zhan2019} found two prominent rotation periods for TIC~289840928, which is in a spatially resolved binary system with TIC~289840926 and both stars are blended in a single TESS pixel. In our re-analysis,
the TESS pixel-level data revealed the primary (TIC~289840928, M4V, 3100\,K) only has a smooth spot modulation with a period of 15.625\,h, while the secondary (TIC~289840926, M6V, 2800\,K) is the actual complex rotator with 2.400\,h. We thus update all corresponding values here.
}

\subsection{Ages and activity corrections}
\revision{\citet{Zhan2019} estimated the ages of all ten \textit{complex rotators} via their probabilistic membership of young stellar associations, using the \texttt{banyan sigma} software \citep{Gagne2018}, with input from the TESS Input Catalog version 8 \citep[TICv8;][]{Stassun2018} and Gaia Data Release 2 \citep[Gaia DR2][]{Gaia2018}}. They found that all ten targets have a high probability of belonging to young associations (see Table~\ref{tab:updated_parameters_of_the_complex_rotators}).\footnote{There seems to be a typo in \citet{Zhan2019}, as TIC~206544316 is a member of Tucana Horologium (not of AB Doradus).}

\revision{We here conduct an independent estimate of their stellar age.}
Young M dwarfs are pre-main-sequence stars and as such have larger radii than their main-sequence counterparts of the same mass. 
Additionally, they have high levels of activity, leading to activity induced radius inflation and temperature suppression \citep{Stassun2012}.
This could be due to strong chromospheric activity, presumably arising from rapid rotation. 
We can thus use the ANU spectra of four targets (Section~\ref{ss:ANU Spectroscopy}) to explore if these stars have \revision{larger radii than expected for the main-sequence.}

For example, TIC 234295610 shows an H$\alpha$ equivalent width of 6.8\,\AA{} in the ANU spectrum. According to the empirical relations from \citet{Stassun2012}, this equivalent width (EW) predicts a radius inflation of 15.6\% and a temperature suppression of 7.1\%. 
Next, we performed a spectral energy distribution (SED) fit to better constrain the apparent radius to ${R_\star^\mathrm{apparent}}=0.415\pm0.029$\,\Rsun{} and the apparent temperature to $T_\mathrm{eff}^\mathrm{apparent}=3075\pm100$\,K (see Fig.~\ref{fig:SED_collage}).
Assuming that this radius and temperature represent the activity-inflated radius and activity-suppressed temperature values, then in the absence of activity we would have 
$R_\star^\mathrm{w/o~activity}\approx0.36$\,\Rsun{} and
$T_\mathrm{eff}^\mathrm{w/o~activity}\approx3300$\,K.

Comparing the corrected values with models for low-mass pre-main-sequence stars \citep{Baraffe2015}, we find that they are fully consistent with a star of mass of 0.20\,\Msun{} and age of 40\,Myr.\footnote{Models from \citet{Baraffe2015} are for stars without activity effects, hence we compare them with our `corrected' values.}
This leads to a stellar mass that is only half of the mass of a main-sequence star with the same radius and effective temperature.
We consider this a strong affirmation of the young age suspected from its young association membership.

We perform the same revision of stellar parameters for all systems. 
First, we perform SED fits to refine all their $T_\mathrm{eff}^\mathrm{apparent}$ and $R_\star^\mathrm{apparent}$ (see Fig.~\ref{fig:SED_collage}). 
Next, for those with ANU spectra (TIC 177309964, TIC 206544316, TIC 234295610, and TIC 425933644), we measure their H$\alpha$ EWs and use them to compute the temperature suppression and radius inflation factors, providing the corresponding values `without activity'. 
For the other six targets, we provide provisional corrections assuming they have H$\alpha$ EWs comparable to the average of the four measured ANU spectra.

\begin{figure}[htbp]
    \centering
    \includegraphics[width=1\columnwidth]{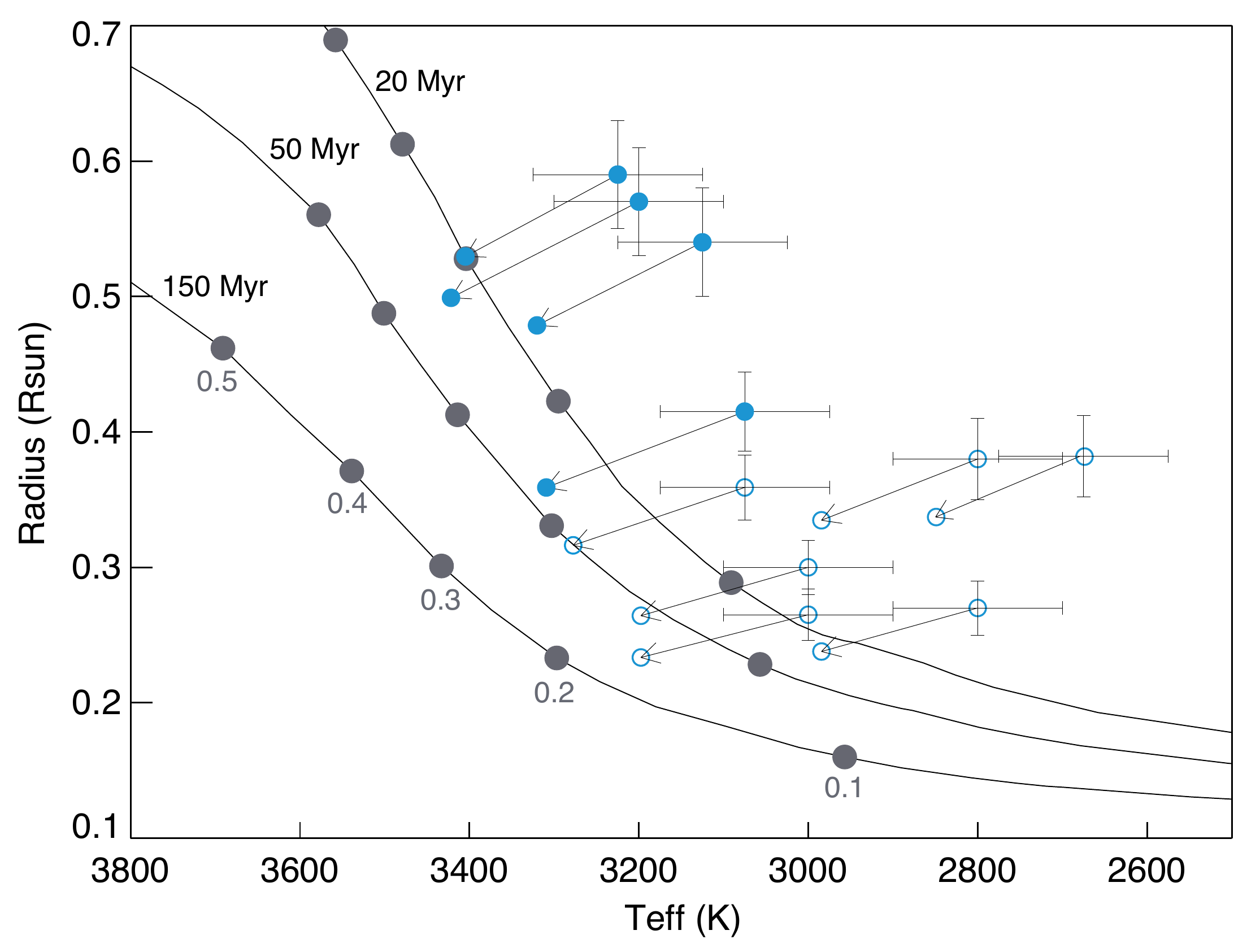}
    \caption{
    Revised stellar parameters of the ten \textit{complex rotators} after correcting for activity-induced radius inflation and temperature suppression.
    Solid curves are isochrones from the \citet{Baraffe2015} models for ages of 20, 50, and 150\,Myr, with gray circles marking masses at 0.1\,\Msun{} increments. 
    Error bars show apparent effective temperatures ($T_\mathrm{eff}$) and radii measured from SED analyses. 
    Arrows show where the stars would fall if they were inactive, i.e., after they are corrected for activity. 
    Filled blue circles mark the four targets with H$\alpha$ measurements from ANU (Section~\ref{ss:ANU Spectroscopy}), whereas empty blue circles mark the remaining six targets for which we provide estimates only.
    The `corrected' values place most of the stars near isochrones consistent with the respective ages estimated from their young association memberships\revision{, suggesting inferred masses of 0.1--0.3\,\Msun{} (Table~\ref{tab:updated_parameters_of_the_complex_rotators})}.
    }
    \label{fig:teff_rad_corr}
\end{figure}

\revision{Fig.~\ref{fig:teff_rad_corr} and  Table~\ref{tab:updated_parameters_of_the_complex_rotators} summarize the revised stellar parameters and their isochrone matches.}
The `corrected' $T_\mathrm{eff}$ and radii place nine of the ten \textit{complex rotators} in the range of $\sim$20 or $\sim$50\,Myr, consistent with the respective ages of their young associations. 
Only TIC~332517282 falls closer to the $\sim$150\,Myr isochrone, consistent with its membership in AB Doradus (150\,Myr), making it our single `oldest' star.


\begin{longrotatetable}
\begin{deluxetable*}{l|cccccccccc}
\tablecaption{Updated parameters of the \textit{complex rotators} from TESS Sectors 1 \& 2.\label{tab:updated_parameters_of_the_complex_rotators}}
\tablewidth{700pt}
\tabletypesize{\footnotesize}
    \tablehead{TIC ID $^1$ & 38820496 & 177309964 & 201789285 & 206544316 & 224283342 & 234295610 & \revision{289840926*} & 332517282 & 425933644 & 425937691}
    \startdata
        RA (deg) $^1$ & 7.19516 & 103.45258 & 33.88868 & 18.41884 & 356.35758 & 357.98325 & \revision{317.629044} & 350.8786 & 3.69942 & 5.36556 \\
        Dec (deg) $^1$ & -67.86237 & -75.70396 & -56.45488 & -59.65974 & -40.33782 & -64.79293 & \revision{-27.18098} & -28.12114 & -60.06352 & -63.85226 \\
        Association $^2$ & Tuc. Hor. & Carina & Tuc. Hor. & Tuc. Hor. & Columba & Tuc. Hor. & $\beta$ Pictoris & AB Doradus & Tuc. Hor. & Tuc. Hor. \\
        \revision{Membership} Probability $^2$ & 99.9\% & 83.2\% & 99.7\% & 100.0\% & 75.0\% & 99.9\% & \revision{99.0\%} & 99.1\% & 99.8\% & 98.8\% \\
        Age (Myr) $^2$ & $45\pm4$ & $45^{+11}_{-7}$ & $45\pm4$ & $45\pm4$ & $42^{+6}_{-4}$ & $45\pm4$ & $24\pm3$ & $149^{+51}_{-10}$ & $45\pm4$ & $45\pm4$ \\
        Distance (pc) $^1$ & $44.06\pm0.12$ & $91.08\pm0.49$ & $45.16\pm0.17$ & $43.07\pm0.10$ & $38.12\pm0.10$ & $48.17\pm0.11$ & $40.33\pm0.18$ & $39.08\pm0.10$ & $44.26\pm0.18$ & - \\
        Rotation Period (h) $^3$ & 15.71 & 10.88 & 3.64 & 7.73 & 21.35 & 18.28 & 2.40 & 9.67 & 11.67 & 4.82 \\
        H$\alpha$ \revision{Equivalent Width (A)} $^4$ & (5.3) & 5.8 & (5.3) & 4.8 & (5.3) & 6.8 & (5.3) & (5.3) & 3.8 & (5.3) \\
        \revision{App. Eff. Temp.,} $T_\mathrm{eff}^\mathrm{app}$ (K) $^5$ & $3000\pm100$ & $3200\pm100$ & $2800\pm100$ & $3100\pm100$ & $3100\pm100$ & $3100\pm100$ & \revision{$2700\pm100$} & $3000\pm100$ & $3200\pm100$ & $2800\pm100$ \\
        \revision{App. Radius,} $R_\star^\mathrm{app}$ ($R_\odot$) $^5$ & $0.30\pm0.02$ & $0.57\pm0.04$ & $0.27\pm0.02$ & $0.54\pm0.04$ & $0.36\pm0.02$ & $0.42\pm0.03$ & \revision{$0.38\pm0.03$} & $0.27\pm0.02$ & $0.59\pm0.04$ & $0.38\pm0.03$ \\
        \revision{Temp. Suppression} $^6$ & (6.2\%) & 6.5\% & (6.2\%) & 6\% & (6.2\%) & 7.1\% & (6.2\%) & (6.2\%) & 5.3\% & (6.2\%) \\
        \revision{Radius Inflation} $^6$ & (13.5\%) & 14\% & (13.5\%) & 13\% & (13.5\%) & 15.6\% & (13.5\%) & (13.5\%) & 11.5\% & (13.5\%) \\
        \revision{Eff. Temp.,} $T_\mathrm{eff}^\mathrm{w/o~act}$ (K) $^6$ & (3200) & $3400\pm100$ & (3000) & $3300\pm100$ & (3300) & $3300\pm100$ & \revision{(2800)} & (3200) & $3400\pm100$ & (3000) \\
        \revision{Radius,} $R_\star^\mathrm{w/o~act}$ ($R_\odot$) $^6$ & (0.26) & $0.50\pm0.04$ & (0.24) & $0.48\pm0.04$ & (0.32) & $0.36\pm0.03$ & \revision{(0.33)} & (0.23) & $0.53\pm0.04$ & (0.33) \\
        \revision{Mass,} $M_\star$ ($M_\odot$) $^7$ & (0.15) & 0.29 & (0.08) & 0.22 & (0.19) & 0.2 & \revision{(0.08)} & (0.16) & 0.29 & (0.08) \\
        \revision{Surface Gravity,} $\log_{10} (g)$ & (4.7) & 4.4 & (4.5) & 4.3 & (4.6) & 4.5 & \revision{(4.2)} & (4.8) & 4.4 & (4.2) \\
        \revision{Co-rotation Radius, $R_\mathrm{cr} (R_\star)$} & \revision{5.6} & \revision{2.9} & \revision{1.9} & \revision{2.2} & \revision{6.2} & \revision{4.9} & \revision{1.0} & \revision{4.6} & \revision{2.9} & \revision{1.6} \\
        TESS Sectors & 1--2 & 1--13 & 2--3 & 1--2 & 2 & 1 & 1 & 2 & 1--2 & 1--2
    \enddata
\end{deluxetable*}
\hspace*{-9.8cm}
\begin{minipage}{8.8in} 
    Tuc. Hor.: Tucana Horlogium;
    $^1$ from TICv8 \citep{Stassun2018}; 
    $^2$ via Banyan Sigma \citep{Gagne2018};
    $^3$ via TESS photometry;
    $^4$ via ANU spectroscopy;
    $^5$ via SED fit;
    $^6$ via \citep{Stassun2012};
    $^7$ via \citep{Baraffe2015} models for a star with the \revision{respective age,} $T_\mathrm{eff}^{w/o~act}$, and $R_\star^{w/o~act}$;
    Values in parentheses are only estimated using the mean H$_\alpha$ equivalent width of the other four targets, and should only be used as an approximate guidance. 
    * TIC~289840926 and TIC~289840928 form a binary system that is blended in a single TESS pixel, but is spatially resolved by other surveys. TIC~289840926 is the complex rotator (see Section~\ref{ss:Revisiting the binary TIC289840928 and TIC289840926}). 
\end{minipage}
\end{longrotatetable}

\section{Flares and sudden brightenings}
\label{s:Flares and sudden brightenings}

The \textit{complex rotators} and \textit{scallop shells} show frequent and large-amplitude flaring, along with other sudden brightenings whose shapes are \revision{distinct} from usual M dwarf flare profiles \citep{Stauffer2017, Stauffer2018, Zhan2019}. 
In particular, brightenings of the entire modulation often appeared right after strong flares, sometimes even followed by changes in the overall morphology\revision{, underlining strong magnetic activity.}

It is still disputed whether flares on stars other than our Sun correlate with the rotational phase and are linked to localized clusters of spots on the stellar surface.
Many previous studies found that superflares were distributed randomly uniform over the rotational phase for main sequence dwarfs \citep{Roettenbacher2018, Doyle2018, Doyle2019, Doyle2020}, and young stars \citep{Vida2016, Feinstein2020}.
\citet{Doyle2018} reason that depending on the viewing geometry, polar spots could be seen at all phases, and their interaction with emerging active regions can thus cause continuously visible flaring.
However, other studies found flares to be more prominent at certain rotational phases, and hence potentially bound to the locations of star spots for the Sun \citep{Zhang2008}, Sun-like stars \citep{Notsu2013}, and the smallest flares on main sequence dwarfs \citep{Roettenbacher2018}.
Hence, a unifying idea is that superflares occur over the entire surface while small flares are tied to spots. 

Here, we utilize the extended coverage by TESS (up to one year for TIC~177309964) to study whether \textit{complex rotators}' flares and other brightenings correlate with the phase of the modulation. 
We searched the light curves in two ways. 
In the first approach, we ran the \texttt{stella} software, a convolutional neural network developed for probabilistic flare detection in TESS 2 min. cadence data \citep{Feinstein2020,stella-code}. 
As the algorithm was trained on a large sample of stars with smooth spot modulation, many of the initial flare candidates were misidentified (often spikes of the \textit{complex rotators}).
By visually vetting, we then \revision{selected} reliable criteria of a probability $\geq$0.9 and amplitude $\geq$5\%, and identify a confirmed sample of at least 67 flares on TIC~177309964.
In the second approach, we independently inspected the entire light curve by eye, and found a total of $\sim$70 confirmed flares, agreeing with the machine learning results.

\begin{figure}[htbp]
    \centering
    \includegraphics[width=0.9\columnwidth]{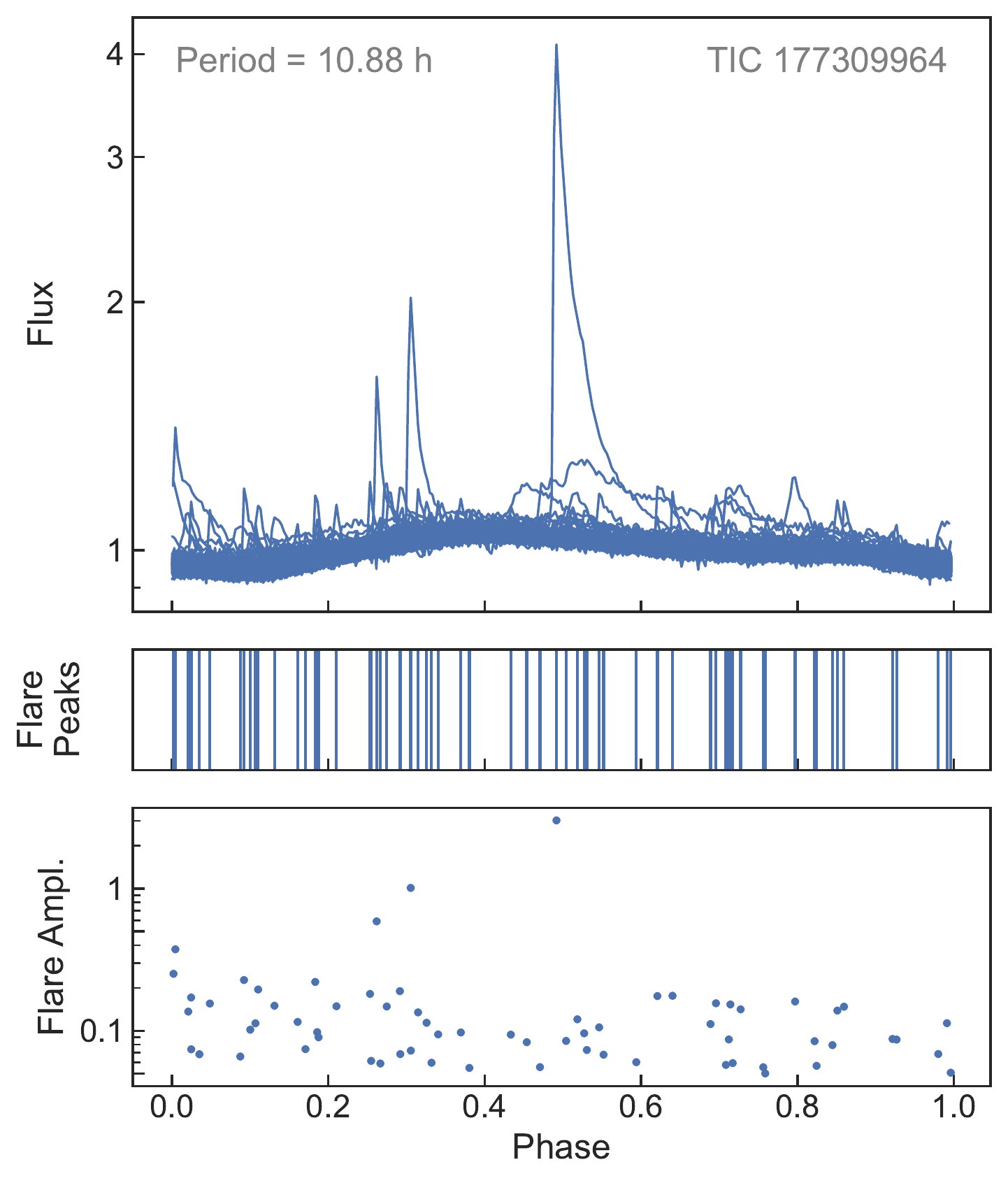}
    \caption{
    Flares on \textit{complex rotators} appear randomly uniform distributed in phase, exemplified by one year of TESS data on TIC~177309964.
    The top panel shows the phase folded light curve with all flares. 
    The middle panel shows the locations of all flare peaks, illustrating the randomly uniform distribution.
    The bottom panel illustrates that there is no clear correlation between flare amplitude and phase.
    }
    \label{fig:flares_TIC177}
\end{figure}

We find that flares on the \textit{complex rotators} are distributed randomly uniform in phase, showing no clear dependency of their location or amplitude (see example of TIC~177309964 in Fig.~\ref{fig:flares_TIC177}). 
There is also no clear correlation between flare amplitudes and the one year time span. 
It is peculiar that the three largest flares (amplitudes of 1.5 to 4 in normalized flux) all occur within two few weeks from one another, but the sample size is too low to rule out mere coincidence.
Most flares are described by the same profiles as their main sequence counterparts, suggesting similar origins and processes driving them.
We also observed somewhat more complicated `outbursts' of flares, which again resemble those of main sequence M dwarfs; these can be explained as superpositions of multiple flare events \citep[e.g.][]{Guenther2020a}.

\begin{figure}[htbp]
    \centering
    \includegraphics[width=0.95\columnwidth]{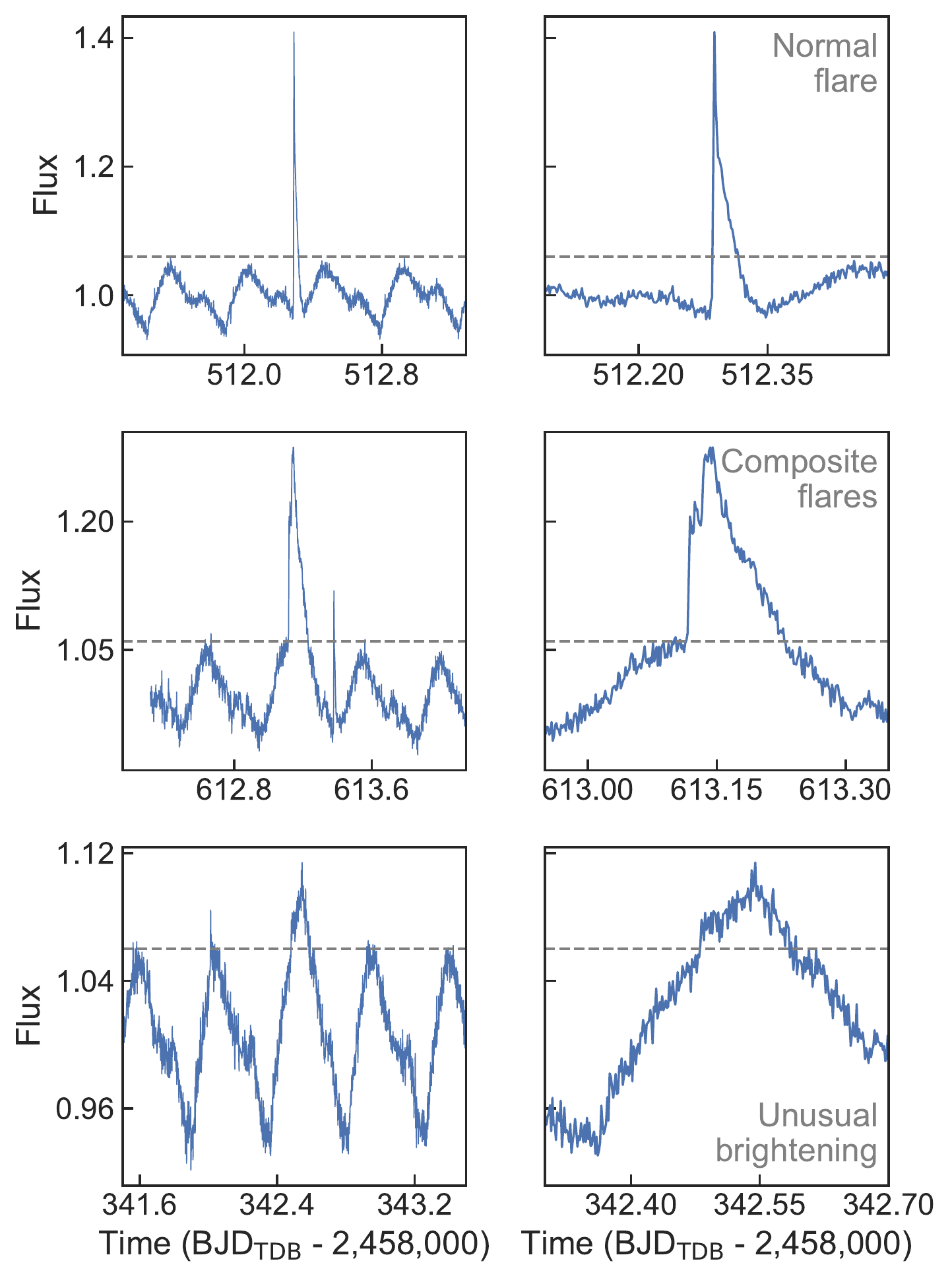}
    \caption{
    Different flare and brightening profiles of the \textit{complex rotators}, exemplified by TIC~177309964. 
    Each row shows a global (2\,day) and local (5\,hr) view of the light curve brightening.
    Horizontal dashed lines show the typical non-flare maximum flux value.
    \textbf{First row:} normal flares are described by the same profiles as main sequence M dwarfs, suggesting similar origins and processes driving them.
    \textbf{Second row:} complicated outbursts of flares also resemble those of main sequence M dwarfs, and can be explained as superpositions of sequential flare events \citep[e.g.][]{Guenther2020a}.
    \textbf{Third row:} unusual brightenings in the \textit{complex rotators} often, but not always, occur after large flares, and show an amplified version of the usual morphology for a short time.
    }
    \label{fig:flares_examples_TIC177}
\end{figure}

Finally, we also find that some sudden brightenings do not resemble typical M dwarf flare profiles (Fig.~\ref{fig:flares_examples_TIC177}); instead, they seem like amplified versions of the complex periodic morphology.
This was also pointed in \citet{Zhan2019} and \citet{Stauffer2017}, often (but not always) following superflares.
On TIC~177309964, these alterations also occur without any preceding flare observation. 
It is possible that they were triggered by a superflare that was (i) not visible in the visual, but would have been classified as such in the UV or X-ray spectrum \citep[e.g.][]{Wolter2007, Loyd2020}, or (ii) was located outside of the visible hemisphere.


\section{Testing the hypotheses}
\label{s:Testing the hypotheses}

The limitations of previous hypotheses leave us with two remaining possibilities for the \textit{complex rotators} and \textit{scallop shells}: (i) the idea of a patchy torus of clouds of gas at the Keplerian co-rotation radius (\textit{co-rotating clouds} hypothesis) and (ii) the idea of spots being periodically occulted behind a spin-orbit-misaligned dust disk (\textit{spots and misaligned disk} hypothesis).
In the following analyses, we hence focus on these two cases.

\subsection{Occurrence rates}
\label{ss:Occurrence rates}

We can estimate the expected yield of \textit{complex rotators} in \TESS{} Sectors 1\,\&\,2 from the \textit{co-rotating clouds} and \textit{spots and misaligned disk} hypotheses (see Fig.~\ref{fig:hypotheses}), respectively, as
\begin{align}
\label{eq:occurrence rates cc}
    N_\mathrm{comp. rot.}^\mathrm{CC} = N_\mathrm{yMrr}  
                                        \cdot P_\mathrm{clouds} 
                                        \cdot P_\mathrm{geom}
\end{align}
and
\begin{align}
\label{eq:occurrence rates ssd}
    N_\mathrm{comp. rot.}^\mathrm{SMD} = N_\mathrm{yMrr}
                                        \cdot P_\mathrm{spots}
                                        \cdot P_\mathrm{disk} 
                                        \cdot P_\mathrm{misal}
                                        \cdot P_\mathrm{geom}.
\end{align}
Here, $N_\mathrm{yMrr}$ is the number of young M dwarfs with rapid rotations ($\lesssim$\,2\,days) which were observed by TESS in short-cadence mode during Sectors 1\,\&\,2.
$P_\mathrm{clouds}$ is the probability that a given star has a cloud of dust/gas orbiting it at the Keplerian co-rotation radius.
$P_\mathrm{spots}$ is the probability that a given star has at least one large spot.
$P_\mathrm{disk}$ is the probability that these stars have a dust disk orbiting them at a few stellar radii.
$P_\mathrm{misal}$ is the probability that a given star shows a spin-orbit misalignment between the stellar rotation axis and the dust disk.
Finally, $P_\mathrm{geom}$ is the geometric probability that an existing structure (cloud or disk) falls in the line of sight between the star and the observer.

\subsubsection{How many young M dwarfs with rapid rotations are out there?}
\label{sss:How many young M dwarfs with rapid rotations are out there?}

\begin{figure}[!htbp]
    \centering
    \includegraphics[width=\columnwidth]{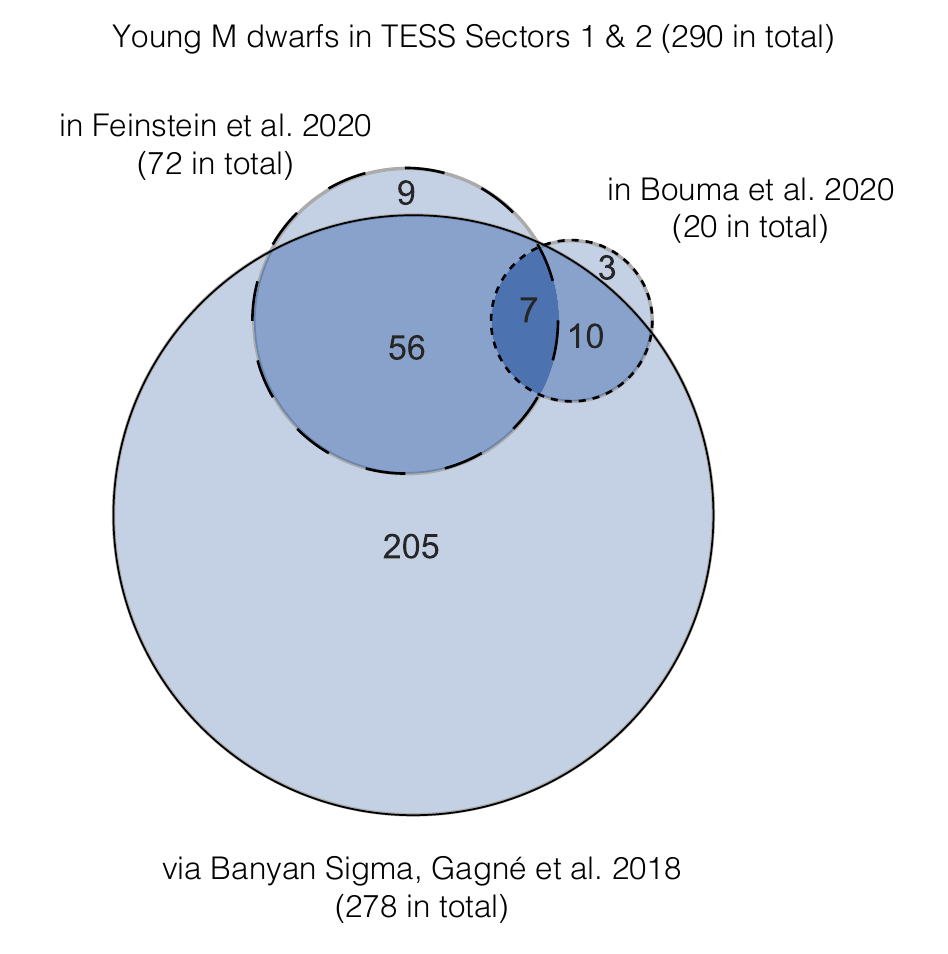}
    \caption{
    Overlap and union of three young star catalogs cross-matched with the TESS short-cadence targets from Sectors 1\,\&\,2. The three catalogs were assembled by \citet{Feinstein2020}, \citet{Bouma2019}, and in this work using \texttt{banyan sigma} \citep{Gagne2018}.
    }
    \label{fig:young_star_catalogs}
\end{figure}

To first order, we can estimate $N_\mathrm{yMrr}$ as the number of stars in known open clusters and associations that have effective temperatures below 3900\,K and radii below 0.6\,\Rsun{}. 
For this, we cross-match the TESS short cadence target lists\footnote{\url{https://tess.mit.edu/observations/target-lists/} (June 24, 2020)} of Sectors 1\,\&\,2 with three young star catalogs: 
\begin{enumerate}
    \item a catalog by \citet{Feinstein2020}, which was assembled through a combination of searching the TESS Guest Investigator proposals and the data from \citet{Faherty2018}. All targets were vetted with the \texttt{banyan sigma} software \citep{Gagne2018} and the ones with $\geq 50$\% probability and available TESS 2\,min data were included.
    \item a catalog by \citet{Bouma2019}\revision{, collecting targets from numerous literature lists, including members of open clusters, 
    moving groups and young associations.} 
    \item a catalog we created by matching all TESS short-cadence targets with the \texttt{banyan sigma} software \citep{Gagne2018}. The algorithm uses a Bayesian model to predict whether a given target is part of one of 27 young associations within 150 parsec. The target is identified by its coordinates, proper motion, parallax, distance, and radial velocity measurements\revision{, which we retrieve from TICv8 and Gaia DR2.}
\end{enumerate}
The results of our cross-matches are shown in Figure~\ref{fig:young_star_catalogs}. We find a total of 290 young M dwarfs from open clusters and young associations. \revision{Notably, their real number might be even higher,} as new clusters and associations are still being discovered. For example, \citet{Gagne2020} just discovered a new young association at 150\,pc, while \citet{Castro-Ginard2020} recently discovered 582 new open clusters in the Galactic disc using Gaia DR2, \revision{increasing their total number} by 45\% (many beyond 150\,pc). Additionally, there might still be unidentified M dwarfs among nearby clusters. We thus consider the number of young M dwarfs in TESS Sectors 1\,\&\,2 as $N_\mathrm{yM} \gtrsim 290$. 

\begin{figure}[!htbp]
    \centering
    \includegraphics[width=\columnwidth]{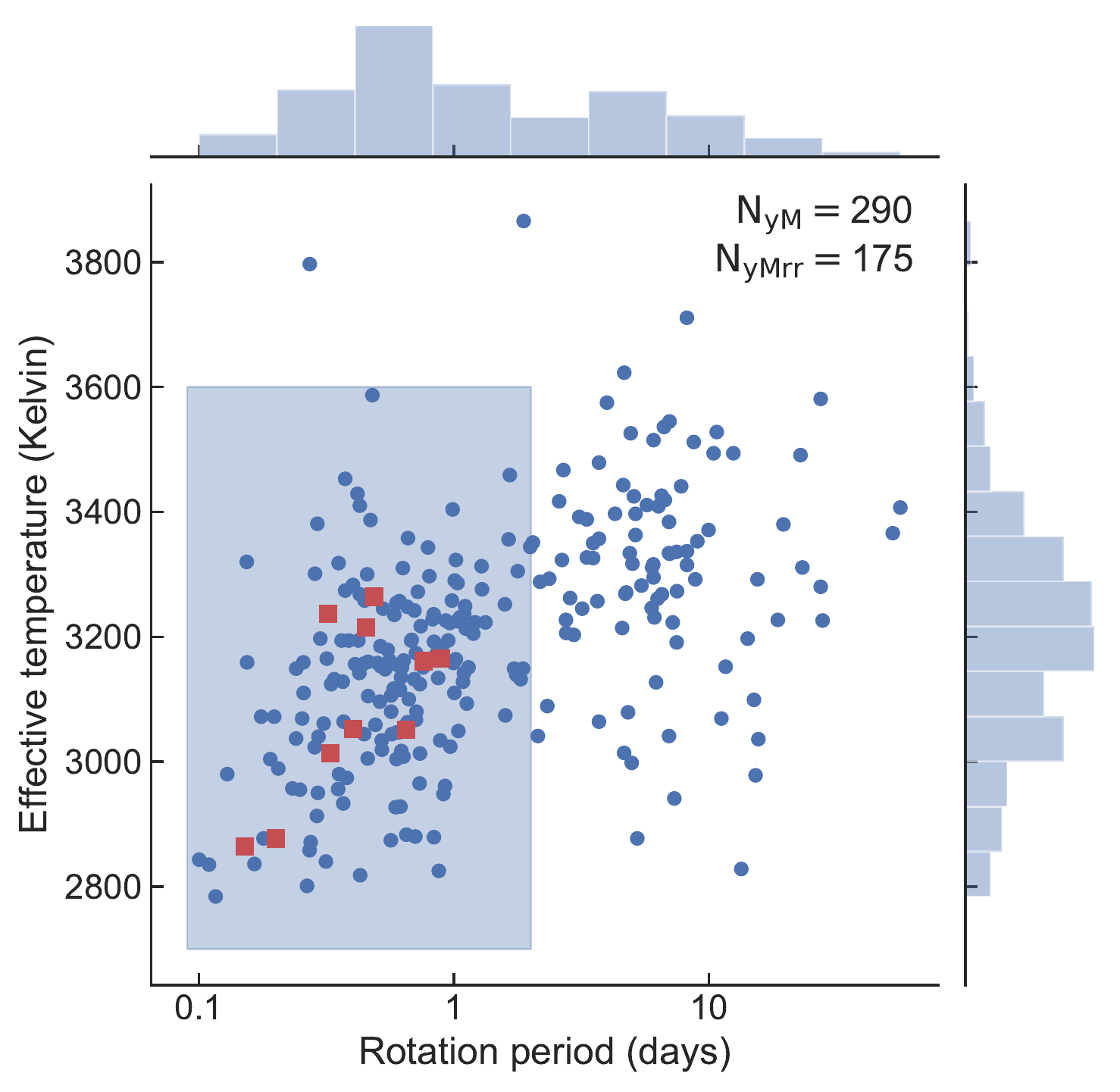}
    \caption{
    Effective temperatures and rotation periods of the 290 young M dwarf stars in TESS short-cadence from Sectors 1\,\&\,2 (blue circles; $\mathrm{N_{yM}}$: number of young M dwarfs). 
    Out of these, 175 targets show rotation periods shorter than 2\,days (blue shaded area; $\mathrm{N_{yMrr}}$: number of young M dwarfs with rapid rotations), comparable to the ten known \textit{complex rotators} from this sample (red squares).
    }
    \label{fig:NyMrr}
\end{figure}

To estimate $N_\mathrm{yMrr}$ from this, we investigate the rotation periods of those 290 M dwarfs using all available TESS data from Sectors 1 through 13. We flag a target as a `young M dwarf with rapid rotation' if we find a rotation period below 2 days using Lomb-Scargle periodograms \citep{Lomb:1976, Scargle:1982}.
We measure rotation periods for 269 out of the 290 targets, with the remaining 21 targets not showing measurable photometric variability (false alarm probability $>$ 0.01).
\revision{This might just be due to low signal-to-noise, or the stars might simply be seen pole-on; both leaves open that they could actually be rapidly rotating.
All measured rotation periods and effective temperatures are shown in Fig.~\ref{fig:NyMrr}, and we conclude that $N_\mathrm{yMrr} \gtrsim $ 175.}


\subsubsection{How many \textit{complex rotators} should we expect?}
\label{sss:How many complex rotators should we expect?}

We here put the finding of $N_\mathrm{yMrr} \gtrsim $ 175 into the perspectives of Eqs.~\ref{eq:occurrence rates cc} and \ref{eq:occurrence rates ssd}. 

\paragraph{\textit{Co-rotating clouds} hypothesis}
For this hypothesis, we also have to account for (i) $P_\mathrm{clouds}$, the probability of clouds of dust/gas being present at the Keplerian co-rotation radius, $R_\mathrm{cr}$, as well as (ii) $P_\mathrm{geom}^\mathrm{CC}$ the geometric alignment probability of an edge-on alignment to the observer.
We can estimate\footnote{
assumes (i) that the clouds are much smaller than the star and (ii) an isotropic distribution of orientations.} 
$P_\mathrm{geom}^\mathrm{CC} \approx (R_\star / R_\mathrm{cr})$, 
with $R_\mathrm{cr}$ derived as:
\begin{equation}
    R_{\mathrm{cr}}=\left(\frac{P_\mathrm{rot}}{2 \pi}\right)^{\frac{2}{3}}\left(G M_{\star}\right)^{\frac{1}{3}},
\end{equation}
with the rotation period $P_\mathrm{rot}$, gravitational constant $G$ and stellar mass $M_\star$.
\revision{
For our targets, this yields $R_\mathrm{cr} \approx 1-6\,R_\star$ (Table~\ref{tab:updated_parameters_of_the_complex_rotators}), which leads to $P_\mathrm{geom}^\mathrm{CC} \approx (R_\star / R_\mathrm{cr}) \approx 0.2-1$.
Putting all pieces together, we can estimate from Eq.~\ref{eq:occurrence rates cc}:
\begin{align}
\begin{split}
    N_\mathrm{comp. rot.}^\mathrm{CC} 
                &= N_\mathrm{yMrr} \cdot P_\mathrm{clouds} \cdot P_\mathrm{geom}^\mathrm{CC} \\
                &\approx 175 \cdot P_\mathrm{clouds} \cdot (0.2-1) \\
                &\approx (35-175) \cdot P_\mathrm{clouds}.
\end{split}
\end{align}
Given that we found 10 \textit{complex rotators} in Sectors 1\,\&\,2, this would only require $\lesssim30$\% of all rapidly rotating young M dwarf to have material clouds trapped at their Keplerian co-rotation radius.
}

\paragraph{\textit{Spots and misaligned disk} hypothesis}
For this hypothesis, we already searched all known young stars in TESS Sectors 1\,\&\,2 for photometric rotation periods and signs of spots, and found that, on average, $P_\mathrm{spots} \sim 1$ (see above).
We estimate the geometric probability $P_\mathrm{geom}^\mathrm{SMD} \approx (R_\star / d)$, where $R_\star$ is the stellar radius and $d \approx 5-15\,R_\star$ is the outer edge (or gap) of a possible disk \citep{Zhan2019}. This leads to $P_\mathrm{geom}^\mathrm{SMD} \approx 0.07 - 0.2$ and:
\begin{align}
\begin{split}
    N_\mathrm{comp. rot.}^\mathrm{SMD} 
            &= N_\mathrm{yMrr} \cdot P_\mathrm{spots} \cdot P_\mathrm{disk}  \cdot P_\mathrm{misal} \cdot P_\mathrm{geom}^\mathrm{SMD}\\
            &\approx 175 \cdot 1 \cdot P_\mathrm{disk} \cdot P_\mathrm{misal} \cdot (0.07-0.2)\\
            &\approx (12-35) \cdot P_\mathrm{disk} \cdot P_\mathrm{misal}
\end{split}
\end{align}
Considering our 10 \textit{complex rotators} in Sectors 1\,\&\,2, this would imply that $P_\mathrm{disk} \cdot P_\mathrm{misal}$ is on the order of $0.3-1$. \revision{Hence, for this hypothesis to hold true, a large fraction of $\gtrsim30$\% of rapidly rotating young M dwarfs would need to have an inner dust disk with a slight spin-orbit misalignment to their rotation axis.}

\subsection{Time dependency}
\label{ss:Time dependency}

\begin{figure}[!htbp]
    \centering
    \includegraphics[width=\columnwidth]{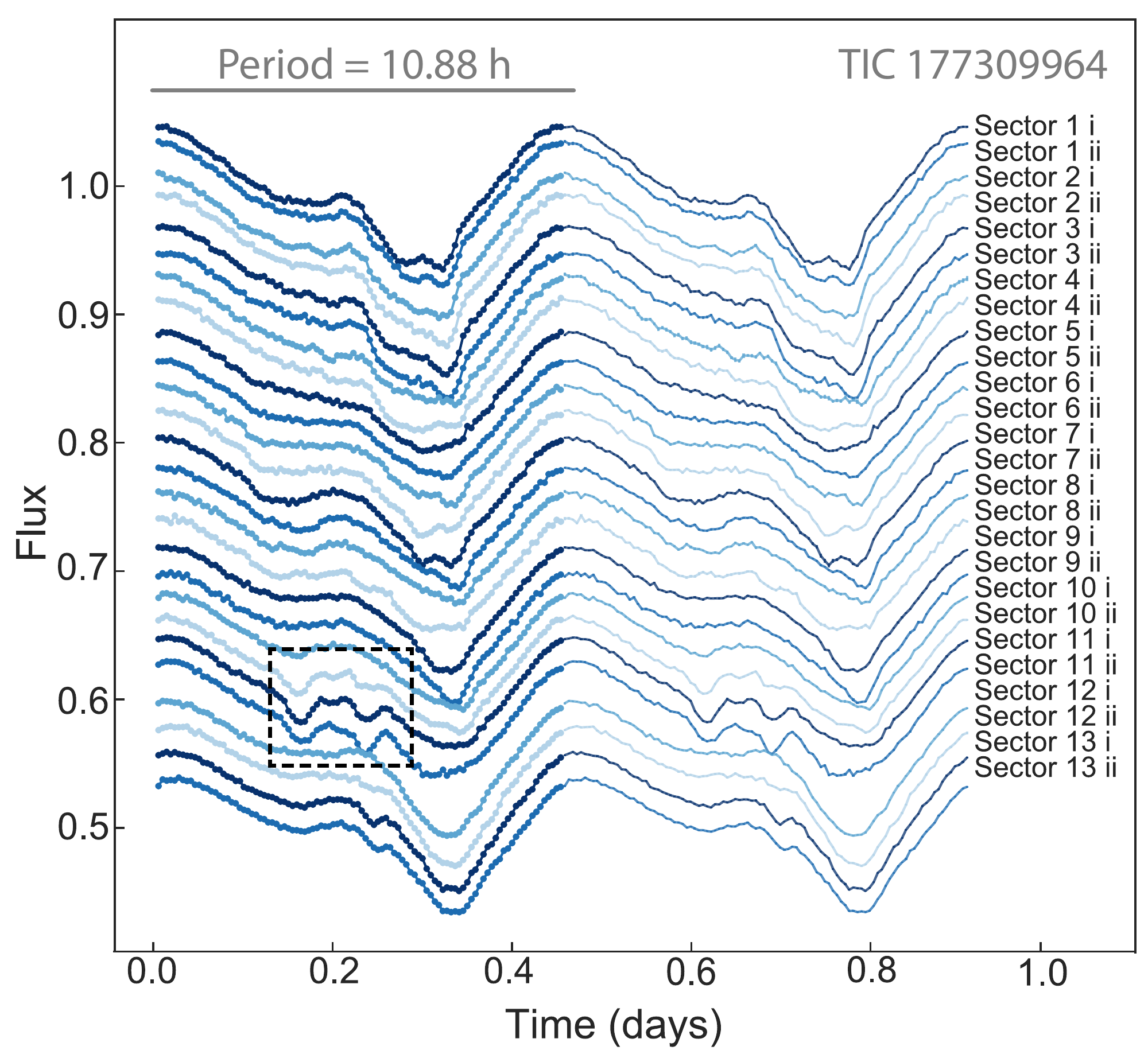}
    \caption{\revision{Example of TIC~177309964, illustrating the stability of major photometric features over year-long timescales (TESS orbits from Sectors 1--13, from July 2018 until July 2019).} Each orbit is phase-folded to show two periods of the modulation and vertically offset for clarity. Flares have been clipped before plotting. The major features of the modulation remain stable over the full year, while minor features appear and disappear over a few weeks (dashed box).}
    \label{fig:longevity}
\end{figure}

The photometric modulations of the \textit{complex rotators} appear stable over timescales of at least one year (Fig.~\ref{fig:longevity} and Fig.~\ref{fig:complex_rotators_multicolor}). 
The \textit{complex rotator} TIC~177309964 fell into TESS' continuous viewing zone and was observed for the consecutive Sectors 1--13, spanning an entire year of data from July 2018 until July 2019 (Fig.~\ref{fig:longevity}). 

We can also see this stability on the examples of TIC~201789285, TIC~206544316, TIC~332517282, and TIC~425933644 (Fig.~\ref{fig:complex_rotators_multicolor}) when combining TESS and SSO photometry. 
The original TESS data were taken in Aug-Oct 2018, while the SSO observations were taken in Nov/Dec 2019, over one year later.
Despite the large time span, the modulation profiles still follow the same pattern and periodicity.
\revision{All major features remain the same, while only some minor evolution of the morphology is evident in the SSO light curves compared to TESS.} In the case of TIC~201789285, it appears that a minor feature has increased in amplitude (near $\mathrm{BJD_{TDB}}$ 2,458,795.70), while another has decreased in amplitude (near $\mathrm{BJD_{TDB}}$ 2,458,795.75).

The stability and longevity of these morphologies is extraordinary.
Considering the \textit{spots and misaligned disk} hypothesis, this is well compatible with the life times of dust disks and the persistence of stellar spots on young M dwarfs.
While stellar spots on most stars only last for a few weeks (e.g., on the Sun they last only for 3--6 rotations \citealt{Gaizauskas1983}), we have examples like GJ 1243, which had remarkably constant spot modulation over 10 years observed with Kepler and TESS \citep{Davenport2020}. Notably, GJ 1243 is a member of a young association with an age of around 30--50\,Myr and, as such, quite comparable to our \textit{complex rotators}.
Furthermore, a 200-day photometric monitoring campaign of the open cluster Blanco 1 ($\sim$115 Myr) with the Next Generation Transit Survey (NGTS) suggests that most young M dwarfs display generally stable spot modulation patterns over this baseline, while F, G and early-K dwarfs show moderate-to-significant evolution in their light curve morphologies \citep{Gillen2020}.

\revision{In contrast}, considering the \textit{co-rotating clouds} hypothesis, such a stability and longevity would seem surprising.
The idea does suggest a rather fine-tuning problem with clouds being confined strictly at the co-rotation radius. Any separating clouds would slowly drift away and slightly alter their orbital period. Even for small drifts, a year-long time-span might lead to noticeable changes in the morphologies. This would lead to a certain amount of material away from co-rotation at any given time, which would blur out the strictly periodic signals over year-long time spans.

\subsection{Color dependency}
\label{ss:Color dependency}

\begin{figure*}[!htbp]
    \centering
    \includegraphics[width=0.85\textwidth]{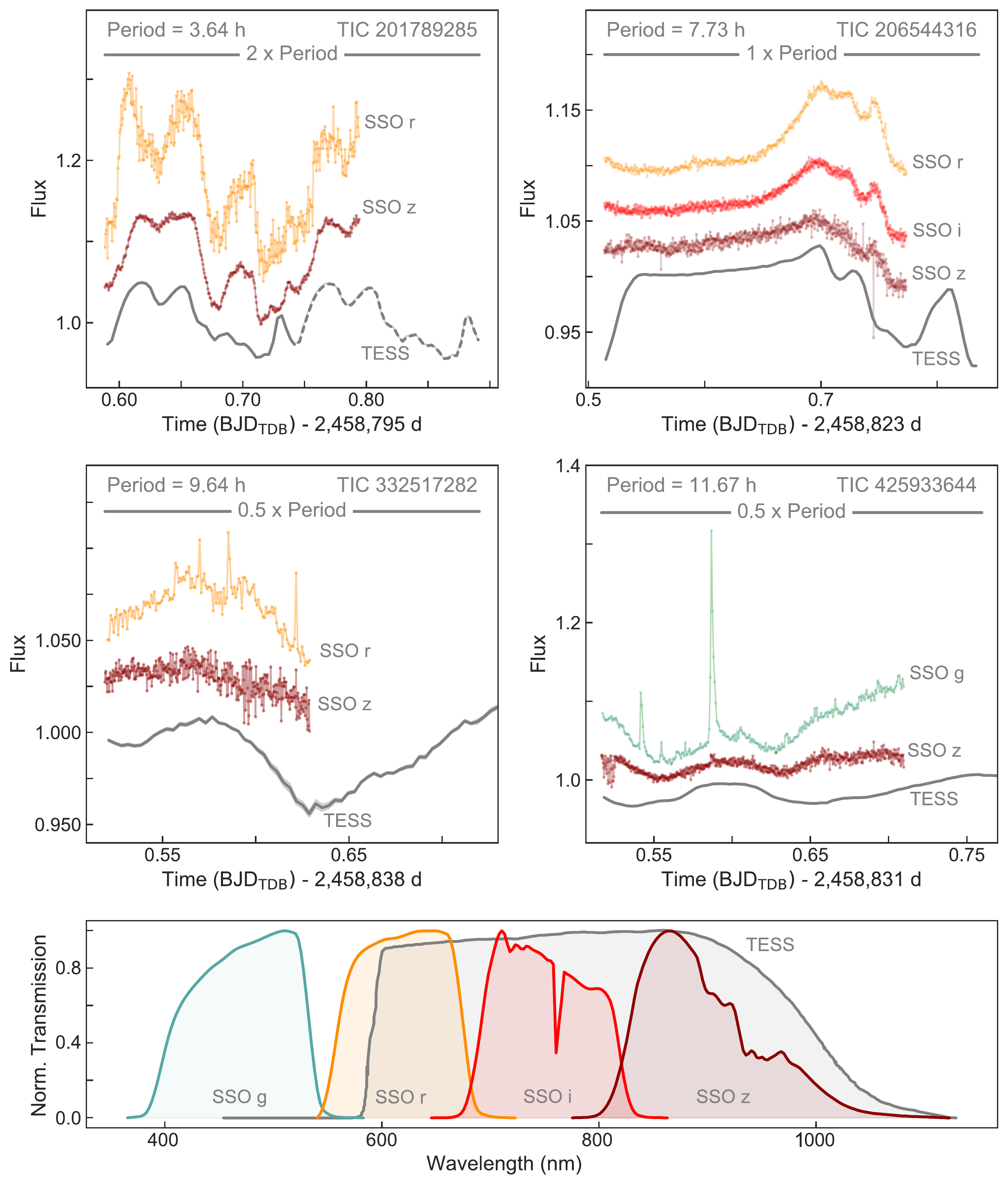}
    \caption{
    \small
    Comparison of multi-color light curves of four of the ten complex rotators: TIC~201789285, TIC~206544316, TIC~332517282, and TIC~425933644.
    For each target, SSO observations were taken simultaneously in at least two of the g', r', i', and z' bandpasses (shown as green, orange, red and dark-red curves, respectively). 
    For TESS observations, flares have been removed, and light curves are averaged over all available Sectors, binned in 5\,min intervals, and slightly shifted in phase to correct for imprecision in the period measurement (shown as dark-gray curves).
    The lower panel compares the normalized transmission functions of all respective bandpasses.
    There is a clear color dependency of the light curve features visible in the simultaneous SSO observations, with features being much more prominent in bluer bandpasses.
    Additionally, the general shape and largest features in the SSO light curves are still comparable to the TESS light curve, even though the SSO data were taken about 1 year later. 
    This suggests that the overall mechanism causing these patterns also causes a color-dependency (e.g., spots, non-gray dust, or pulsations), and that it is stable over long times.}
    \label{fig:complex_rotators_multicolor}
\end{figure*}

We obtained a total of nine telescope nights worth of SSO observations (Section~\ref{ss:SPECULOOS Southern Observatory Photometry}) to capture four of the \textit{complex rotators} in simultaneous multi-color bandpasses.
We compare all these light curves with phase-folded TESS observations (taken one year earlier) in Fig.~\ref{fig:complex_rotators_multicolor}.
Evidently, the sharp-peaked features are more prominent in bluer bandpasses, and less expressed in the reddest bandpasses.
This matches the expectations from both the \textit{co-rotating clouds} and \textit{spots and misaligned disk} hypotheses:
(i) for the \textit{co-rotating clouds} hypothesis, the material's extinction would have to be stronger in the blue, leading to deeper features. 
(ii) for the \textit{spots and misaligned disk} hypothesis, the contrast between the stellar surface ($\sim$3000\,K) and a cool spot is stronger in bluer wavelengths. The disk material could be a gray absorber or could have a color dependency, which would add a secondary effect.

Combining TESS data with SSO r, i and g band observations, our total data span more than one year. We find that the same features are still present in the data at the predicted phases \revision{(accounting for uncertainties in the period estimation)}. 
There is no doubt that the modulation is still the same, and thus stable and long-lived over year-long time spans.

\section{Discussion}
\label{s:Discussion}

\subsection{Could spots and pulsations be another hypothesis?}
\label{ss:Could spots and pulsations be another hypothesis?}

Pulsations on low mass stars have long been theoretically explored and predicted \citep[e.g.,][]{Gabriel1964,Noels1974,Palla2005,Rodriguez-Lopez2012,Rodriguez-Lopez2014,Rodriguez-Lopez2019} but, despite observational efforts, have so far proven elusive. %
In theory, fully non-adiabatic models of M dwarfs suggest that they could excite (i) radial modes, (ii) low-order, low-degree non-radial modes, and (iii) solar-like oscillations \citep{Rodriguez-Lopez2012,Rodriguez-Lopez2014}.
This requires the models to be completely convective or have large convective envelopes. This would match the spectral types of all \textit{complex rotators} and \textit{scallop shells}, placing the stars beyond the fully convective limit.
Periods of the pulsations are predicted to range from 20\,min to 3\,h, again agreeing with the typical time scales we see for the sharp-peaked features of our targets. 

\revision{However, the amplitudes of M dwarf pulsations are expected to be in the range of 1\,ppm to 1\,ppt, while our targets typically show amplitudes of several percent.}
The only bypass to this caveat could be a superposition of the effects from spots (creating the smooth, large-amplitude modulations) and synchronous pulsations (creating the sharp-peaked, lower-amplitude features). 
Yet, this would require spots and pulsations to be synchronized, which seems implausible.

\citet{Rodriguez-Lopez2014} identified the theoretical instability strips of M dwarfs, which resulted in two islands of `instability' in the parameter space. Stars falling into one of these islands would, in theory, be capable of showing pulsations.
With the revised and activity-corrected stellar parameters (Section~\ref{s:Revising the stellar parameters}), some of the stars fall near the lower island, yet remain outside of it. Again, this makes pulsations seem implausible.

\subsection{The pros and cons of various hypotheses: a summary}
\label{ss:The pros and cons of various hypotheses: a summary}

\begin{sidewaysfigure*}
\vspace{1.25in}
    \centering
    \includegraphics[width=\textwidth]{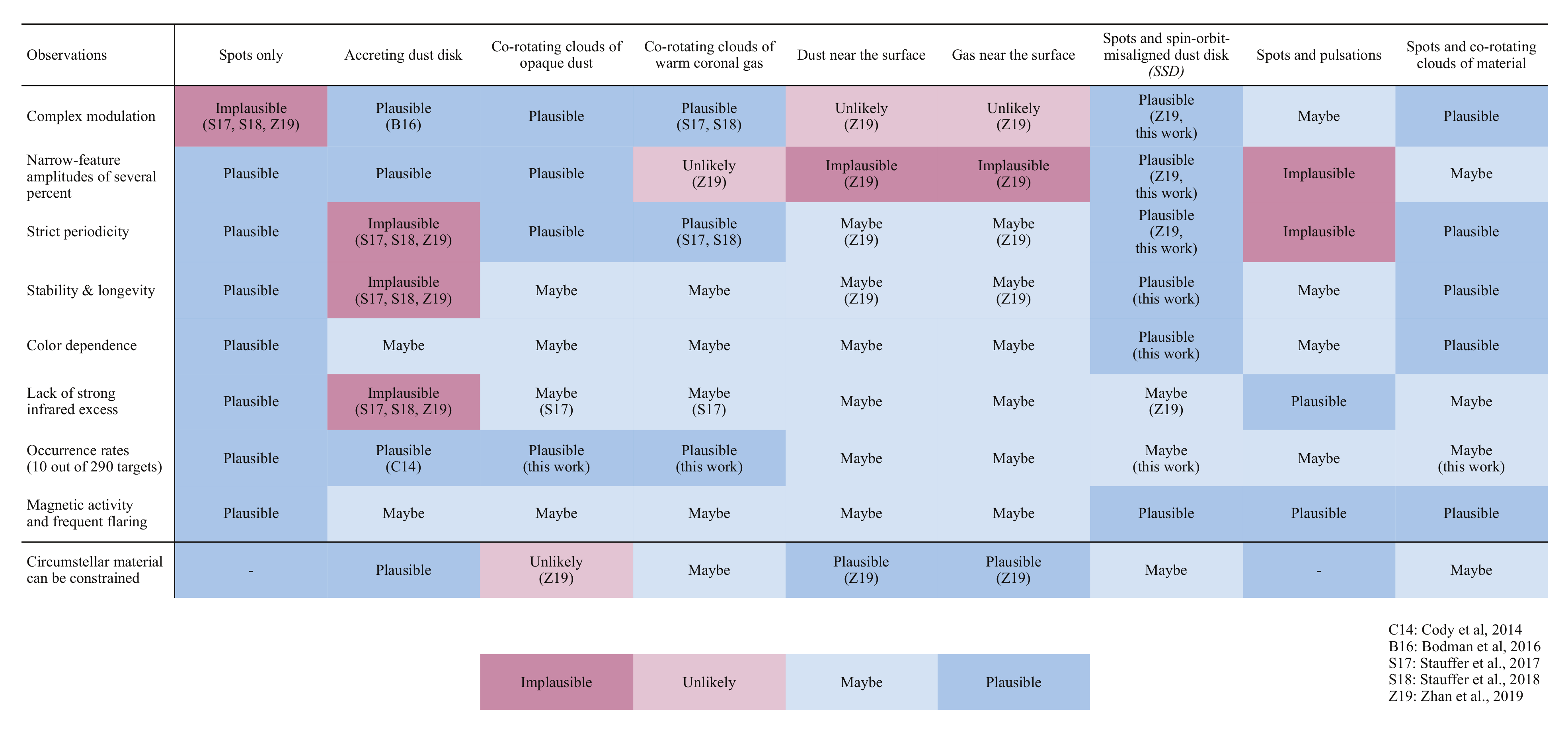}
    \caption{Overview of hypotheses discussed throughout this paper for \textit{complex rotators} (and \textit{scallop shells}). All hypotheses are contrasted against observational constraints.}
    \label{fig:overview}
\end{sidewaysfigure*}

We here briefly summarize the pros and cons of the various hypotheses introduced and scrutinized throughout this paper (see Sections~\ref{s:Introduction} and \ref{ss:Could spots and pulsations be another hypothesis?} for short overviews). 
Fig.~\ref{fig:overview} additionally contrasts all observations with all hypotheses, highlighting which aspects can and cannot be explained through a given hypothesis.

\paragraph{Spots only}
\textit{Pro:} Spot modulations can be strictly periodic and stable over many years, even in the presence of differential rotation \citep[e.g.,][]{Davenport2020}. The temperature difference between the surface and the spots causes a color dependence, and spots would not cause any infrared excess. Most young stars are spotted and are often accompanied by strong signs of magnetic activity, such as flaring.
\textit{Con:} Spots alone cannot explain the sharp-peaked features \citep{Zhan2019}. 
However, spots could still play a major role in combination with other factors (e.g, pulsation or circumstellar material; Sections~\ref{ss:Could spots and pulsations be another hypothesis?} and \ref{ss:Towards a unified hypothesis}).

\paragraph{Accreting dust disk}
\textit{Pro:} Accreting dust disks can lead to the morphologies for \textit{dipper/burster} stars and occur frequently enough. They might show color dependency depending on the absorption and scattering properties of the material.
\textit{Con:} Accretion is a rather stochastic process, and thus neither strictly periodic nor stable. The \textit{dippers} and \textit{bursters} also show strong infrared excess due to the large disks.

\paragraph{\textit{Co-rotating clouds} of material}
\textit{Pro:} Clouds of material at the Keplerian co-rotation radius could qualitatively explain sharp features and strict periodicity \citep{Stauffer2017,Stauffer2018}. Depending on the material, a color dependency is possible, and small enough clouds would cause no infrared excess. As young stars are often surrounded by material, they could also occur at high enough rates.
\textit{Con:} 
If the material is gas, the absorption would likely not be able to explain percentage-scale amplitudes \citep{Zhan2019}.
If the material is dust, these clouds are likely not stable at the required distances ($d/R_\star \gtrsim 3$; \citealt{Zhan2019}).
Another challenge might be the stability and longevity of the morphology over year-long time spans, i.e., over hundreds of orbital periods. With some parts of the clouds slowly drifting away from co-rotation\revision{, the signals would blur out and evolve, which does not seem to be the case.}

\paragraph{Material trapped near the surface}
\textit{Pro:} 
Material trapped in the magnetic field and bound to the stellar rotation would remain strictly periodic, and could survive over many years near the stellar surface ($d/R_\star \sim 1$; \citealt{Zhan2019}).
Depending on the material's properties, a color dependence is possible, and in small amounts it might not cause any infrared excess.
\textit{Con:} 
Any material that close in cannot explain the sharp-peaked, percentage-scale amplitudes of the modulation but would instead produce a rather smooth variation similar to spots; this can only be explained by material at larger distances ($d/R_\star \gtrsim 3$; \citealt{Zhan2019}).

\paragraph{Spots and spin-orbit-misaligned dust disks}
\textit{Pro:}
The patterns can be strictly periodic and stable over many years. Spots induce a color dependency, and disk material might add to this effect. There are enough young and rapidly rotating M dwarfs in the sample to explain the high occurrence rates (with caveats, see below). Lastly, the spots are a sign of magnetic activity, and agree with the frequent flaring found on the \textit{complex rotators} and \textit{scallop shells}.
\textit{Con:} the scenario would require most young M dwarfs to have close-in dust disks with spin-orbit misalignments. There is no obvious formation mechanism that would explain this behavior. 
Also, the one M dwarf for which we have disk and rotation measurements, Au Mic, does appear co-planar.
However, the misalignment does not need not to be very large. A 10$^\circ{}$ obliquity between the spin and magnetic axes of T Tauri stars is reasonable, based on Zeeman studies and recent work by \citet{McGinnis2020}. If the disks are confined by the magnetic field, this slight misalignment could already be enough to mitigate this caveat and cause the observed morphologies. A potential driver for such misalignment might be perturbations from nearby passing stars, either dynamically or through radiation pressure \citep[e.g.][]{Rosotti2014}.


\paragraph{Spots and pulsations}
\textit{Pro:} 
In theory (qualitatively), a superposition of spots and pulsations could lead to a smooth large-amplitude trend (due to spots) superposed by a sharp-peaked small-amplitude pattern (due to pulsations). Both features can be stable over long times, show color dependencies, a lack of infrared excess, and frequent flaring due to their activity.
\textit{Con:} 
This would require a strict synchronization between rotation and pulsation, which seems implausible. Further, the stars do not lie within any of the theoretical instability islands.

\subsection{Towards a unified hypothesis: could it be spots and co-rotating clouds of material?}
\label{ss:Towards a unified hypothesis}

So far, none of the hypotheses stand out as a definite answer, and each come with limitations. 
The two most promising hypotheses have their own caveats. 
On the one hand, the \textit{co-rotating clouds} hypothesis, with clouds of gas at the Keplerian co-rotation radius could remain stable and allow the required viewing angle, but gas likely cannot explain the large, percentage-scale absorption features.
On the other hand, the \textit{spots and misaligned disk} hypothesis could tick all boxes, but has the implicit requirement that a large fraction of rapidly rotating young M dwarfs must have misaligned disks. 
That said, if the disks are confined by the magnetic field, a small misalignment of $\sim$10$^\circ{}$ seems not to be uncommon \citep{McGinnis2020} and could have been induced by nearby passing stars \citep[e.g.][]{Rosotti2014}.

Rapidly rotating young M dwarfs are known to be magnetically active and generally show high spot coverage rates.
Spots alone were ruled out easily, but what was not explored so far is: \textit{how much} can spots explain?

The superposition of smooth, large-amplitude spot modulations and sharp, sudden features from transits of co-rotating clouds of gas could represent a unified hypothesis (\textit{Spots and co-rotating clouds}). 
\revision{This explanation could expand the \textit{co-rotating clouds} hypothesis by requiring only a minimal amount of circumstellar material to cause the overall morphology, making gas clouds a more plausible candidate. It would also mitigate the occurrence rate caveats which challenge the \textit{spots and misaligned disk} hypothesis.}

\subsection{Toy models}
\label{ss:Toy models}

We developed simplified forward models for the three most promising ideas, the (i) \textit{co-rotating clouds}, (ii) \textit{spots and misaligned disk}, and (iii) \textit{spots and co-rotating clouds} hypotheses. 
We took the TESS light curve of TIC~201789285 as an example, and tried to imitate its morphology as closely as possible while keeping the models simple, using a hybrid of statistical inference and manual parameter selection.
We model the star as a fine grid in spherical coordinates. The rotation axis of the star is left as a free parameter, and a quadratic limb darkening effect is applied to each cell based on its orientation relative to the observer. 

For the \textit{spots and misaligned disk} hypothesis,
\revision{we model spots with} four parameters: two angles describing the location on the star, its size, and its temperature.
The disk is parametrized by its inner and outer radius, inclination, and opacity.
The observed flux is computed for each grid cell by integrating the Planck function across the TESS bandpass, accounting for geometric effects, spots and the disk.
We then try to mimic the TESS light curve of TIC~201789285 by choosing a simple model with three cold spots and a fully opaque disk, optimizing their parameters using nested sampling via \texttt{dynesty} \citep{Speagle2020}.

For the \textit{co-rotating clouds} and \textit{spots and co-rotating clouds} hypotheses, \revision{we model the torus of clouds as a series of orbiting spheres whose orbital period matches the rotational period of the star.
For this, we used the \texttt{processing} package (\url{https://processing.org/}) to draw the model and count the flux in each pixel.} 
We then manually evaluated different scenarios of spots and sparse to dense tori of clouds.

We find that all three toy models can replicate the typical morphology of complex rotators (Fig.~\ref{fig:toy_models}). 
Additionally, spots as drivers for an underlying large-amplitude modulation can explain a large portion of the signal, easing the constraints on circumstellar material.

\subsection{What can TESS short-cadence do for us?}
\label{ss:What can TESS short-cadence do for us?}

The TESS \textit{complex rotators} were all found in short-cadence (2\,min) observations.
In contrast, the K2 \textit{scallop shells} were discovered using 30\,min cadence.
However, the longer cadence means that the data are limited by the Nyquist frequency, and numerous harmonics will be missed in the frequency spectrum (see Fig.~\ref{fig:morphology_classes}).
We tested whether we would have discovered the \textit{complex rotators} in the same way from TESS long-cadence (30\,min) observations, by extracting light curves directly from the full frame images (FFI).
For this, we used a TESS FFI photometry pipeline based on the one developed by \citet{Pal2012}. It is designed to extract photometry of faint stars in crowded fields (TESS-mag$>$15) by combining difference and aperture photometry. We selected nearby stars from the Gaia DR2 catalog that could contaminate the target's light curve and applied a principle component analysis (PCA) detrending. 

\begin{figure}[!htbp]
    \centering
    \includegraphics[width=\columnwidth]{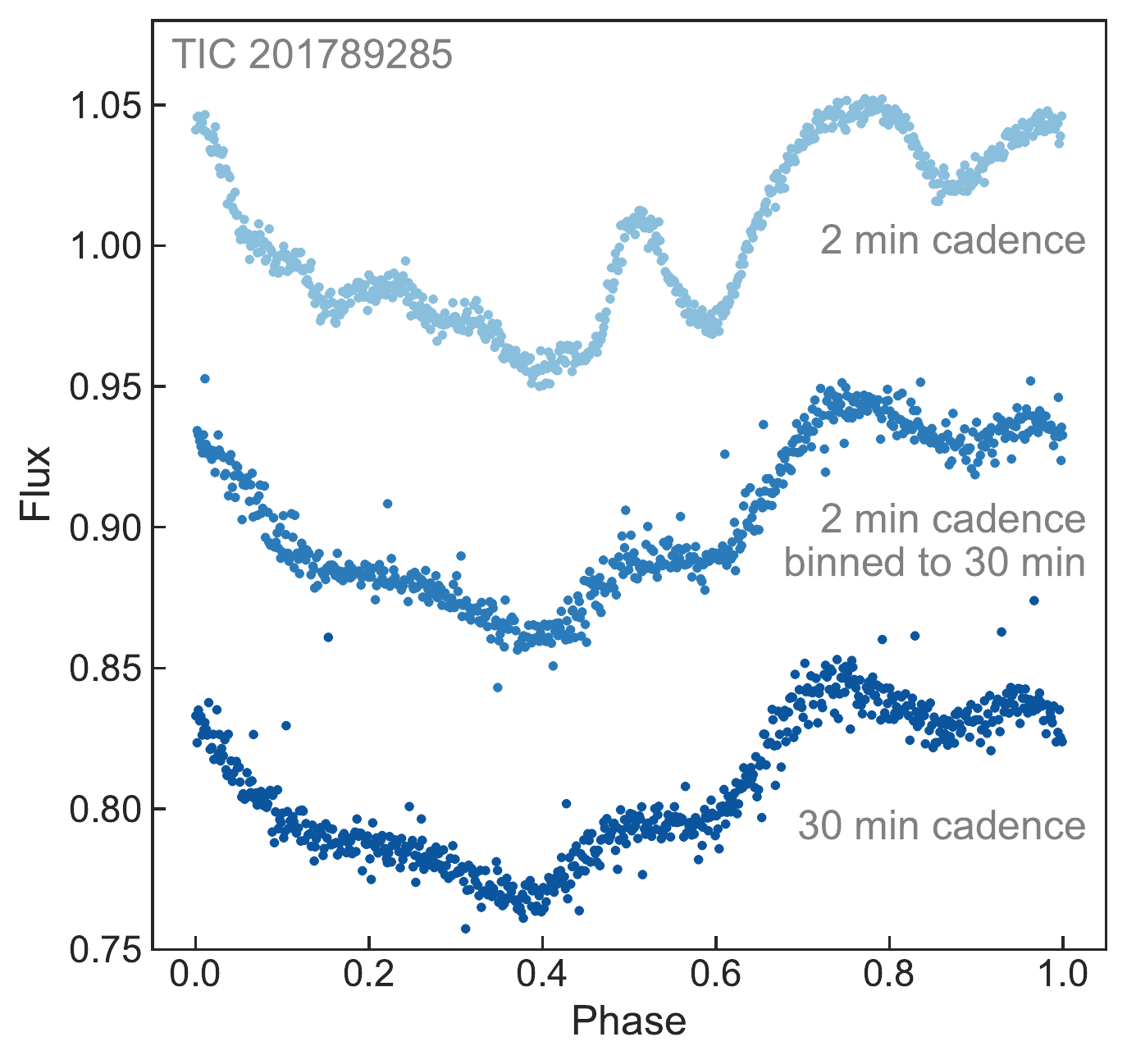}
    \caption{
    Comparison of TESS 2\,min and 30\,min light curves on the example of TIC~201789285, showing that the complex rotators' sharp-peaked features are blurred in long-cadence data. TESS' 2\,min and upcoming 20\,s observing cadences are hence ideal for detecting more of these targets.
    }
    \label{fig:FFI_lightcurve}
\end{figure}

We find that the \textit{complex rotators} morphologies are most apparent in the 2\,min cadence data, and that many might have been missed due to their sharp-peaked features being blurred out in 30\,min cadence data (Fig.~\ref{fig:FFI_lightcurve}).
The 2\,min data will thus be the best source to search for more \textit{complex rotators} throughout Sectors 1--23.
With the start of the TESS extended mission, 20\,s cadence data will be enabled. Additionally, both the original TESS sectors as well as the K2 fields will be revisited. Re-observing \textit{complex rotators} and \textit{scallop shells} with such short cadence could greatly increase our sensitivity to the sharpest features.

\clearpage
\subsection{Implications for exoplanet systems}
\label{ss:Implications for exoplanet systems}

We know thanks to Kepler and other missions that, in average, each early- to mid-M dwarf has at least one small exoplanet \citep{Dressing2015}.
All \textit{complex rotators} and \textit{scallop shells} are young mid-M dwarfs (most around 10 -- 45\,Myr), raising the question of how the effects causing their morphology might impact young, recently formed exoplanets.
After all, for the \textit{co-rotating clouds}, \textit{spots and misaligned disk}, and \textit{spots and co-rotating clouds} hypotheses, occurrence rates suggest that most young M dwarfs would go through the same processes, we just cannot see their morphologies (Section~\ref{ss:Occurrence rates}).

At these ages, most processes driving planet formation have likely been concluded. 
Terrestrial planet formation is thought to be completed after at most 10--30\,Myr, regardless of the driving processes \citep[e.g.,][]{Chambers2010}.
Particularly, gas giant planets around mid- to late M dwarfs are rare; even if they formed around any \textit{complex rotators}, they require a substantial amount of gas in the proto-planetary disk to be present at the late stage of their formation, yielding formation time spans of less than 10\,Myr (the maximal lifetime of gas discs; e.g., \citealt{DAgnelo2010}).

The material in question for causing the complex morphologies, however, is likely much closer to the star (near 3--10 stellar radii) than any forming or migrating exoplanets. 
Exoplanet systems hosted by mid-M dwarfs are widely studied, and if the effects at play for \textit{complex rotators} are indeed ubiquitous at early ages, they seem to not have caused a strong effect on their planets.

If the \textit{spots and misaligned disk} hypothesis were correct, more than half of all young M dwarfs would have a slight spin-orbit misalignment between their rotation axis and remnant dust disk. 
An interesting question is whether this would imply the proto-planetary disk to also be misaligned. However, there are currently no strong signs that a large fraction of exoplanets forming around M dwarfs have spin-orbit misalignments. Limited data suggest there to be a wide variety. For example, Au Mic\,b is proven to be aligned \citep[e.g,][]{Hirano2020a}, the TRAPPIST-1 system might be slightly misaligned \citep[][]{Hirano2020b} and GJ 436\,b is inclined \citep{Knutson2011, Bourrier2017}.

\section{Conclusion}
\label{s:Conclusion}

Recently, at least one new distinct morphology class of young stars has been discovered in white-light photometry from Kepler/K2 and TESS \citep{Stauffer2016a,Stauffer2017,Zhan2019}, adding to the seven classes established by \citet{Cody2014} (which include \textit{dippers} and \textit{bursters}). Here, we added three puzzle pieces to unveil the physical nature of these \textit{complex rotators} and probe whether several hypotheses could hold true given these new observational constraints. 

The tested hypotheses include spots only, accreting dust disks, co-rotating clouds of material, magnetically constrained material, spots and a spin-orbit-misaligned disk, and spots and pulsations. 
We particularly focused on the \textit{co-rotating clouds} and \textit{spots and misaligned disk} hypotheses, as others are ruled out more easily.

First, we investigated if their occurrence rates make sense in the light of the total number of rapidly rotating young M dwarfs in the given field of view. We find that TESS Sectors 1\,\&\,2 harbor at least 175 young M dwarfs with rotation periods below 2 days, rendering the finding of 10 \textit{complex rotators} plausible.
However, for both hypotheses this comes with a caveat.
For the \textit{co-rotating clouds} hypothesis to work, this would mean that almost every such star must have clouds of dust/gas trapped at the Keplerian co-rotation radius.
For the \textit{spots and misaligned disk} hypothesis to work, this would imply that a large fraction of these stars must have an inner disk and show a spin-orbit-misalignment. 
If the latter held true, it could have consequences for exoplanet systems around mid-to-late M dwarfs, which might have formed under the same conditions.

Second, we studied the longevity of these features over one year, and find that they remain remarkably stable over these time spans. While the major features of the \textit{complex rotators} remain unchanged, we find evidence for additional small features building up and decaying over a few weeks. 
In the \textit{Co-rotating Clouds hypothesis}, this would imply subtle changes in the dust/gas cloud structures.
In the \textit{Spots and Misaligned Disks hypothesis}, this can very likely be caused by smaller spots appearing and disappearing, major spots changing size, or spots wandering along the surface.

Third, we probe the color-dependency of the \textit{complex rotators} photometric features. We indeed find the expected behavior predicted by both the \textit{co-rotating clouds} and \textit{spots and misaligned disk} hypotheses. 
The features are more pronounced in bluer wavelengths, which could be explained by either chromatic absorption by the circumstellar material or the smaller spot-to-star brightness contrast in the red / infrared.

All new clues to the case -- occurrence rates, longevity and color-dependency -- could in principle match any of the hypotheses shown in Fig.~\ref{fig:hypotheses}. It is well possible that the truth will lie somewhere between these hypotheses.
Rapidly rotating young M dwarfs are known to be magnetically active, so the final answer will likely have contributions from both spots and circumstellar material, leading to their complex photometric morphologies.

\section*{Acknowledgments}
We thank John Stauffer and Andrew Collier-Cameron for fruitful discussions about complex rotators and scallops, John Bredall and Benjamin J. Shappee for providing their custom detrended light curve of the dipper star RY Lup, Gerald Handler for his insights on pulsation, \revision{and Victor See for discussions about M dwarf activity.}

Funding for the \TESS{} mission is provided by NASA’s Science Mission directorate. 
Resources supporting this work were provided by the NASA High-End Computing (HEC) Program through the NASA Advanced Supercomputing (NAS) Division at Ames Research Center for the production of the SPOC data products.
This paper includes data collected by the \TESS{} mission, which are publicly available from the Mikulski Archive for Space Telescopes (MAST). STScI is operated by the Association of Universities for Research in Astronomy, Inc. under NASA contract NAS 5-26555.
The research leading to these results has received funding from the European Research Council under the European Union's Seventh Framework Programme (FP/2007-2013) ERC Grant Agreement n$^\circ$ 336480, from the European Union's Horizon 2020 research and innovation programme (grant agreement n${^\circ}$ 803193/BEBOP), from the ARC grant for Concerted Research Actions financed by the Wallonia-Brussels Federation, from the Balzan Prize Foundation, from F.R.S-FNRS (Research Project ID T010920F), from the Simons foundation, from the MERAC foundation, and from STFC, under grant number ST/S00193X/1.
This work has made use of data from the European Space Agency (ESA) mission Gaia (\url{https://www.cosmos.esa.int/ gaia}), processed by the Gaia Data Processing and Analysis Consortium (DPAC, \url{https://www.cosmos.esa.int/web/gaia/ dpac/consortium}). Funding for the DPAC has been provided by national institutions, in particular the institutions participating in the Gaia Multilateral Agreement. 

M.N.G. acknowledges support from MIT's Kavli Institute as a Juan Carlos Torres Fellow \revision{and from the European Space Agency (ESA) as an ESA Research Fellow}.
K.O. and B.S. acknowledge support from the Hungarian National Research, Development and Innovation Office grant OTKA K131508. 
B.S. is supported by the \'UNKP-19-3 New National Excellence Program of the Ministry for Innovation and Technology.
J.N.W.\ and B.V.R.\ thank the Heising-Simons Foundation for support.
A.D.F. acknowledges the support from the National Science Foundation Graduate Research Fellowship Program under Grant No. (DGE-1746045). Any opinions, findings, and conclusions or recommendations expressed in this material are those of the author(s) and do not necessarily reflect the views of the National Science Foundation. 
E.G. gratefully acknowledges support from the David and Claudia Harding Foundation in the form of a Winton Exoplanet Fellowship.
M.G. is F.R.S.-FNRS Senior Research Associate.
B.-O.D. acknowledges support from the Swiss National Science Foundation (PP00P2-163967).
J.M.D.K.\ gratefully acknowledges funding from the Deutsche Forschungsgemeinschaft (DFG, German Research Foundation) through an Emmy Noether Research Group (grant number KR4801/1-1), the DFG Sachbeihilfe (grant number KR4801/2-1), and the SFB 881 “The Milky Way System” (subproject B2), as well as from the European Research Council (ERC) under the European Union's Horizon 2020 research and innovation programme via the ERC Starting Grant MUSTANG (grant agreement number 714907).

\textbf{Facilities:} TESS, SSO, ANU

\textbf{Software:} 
\texttt{numpy} (\citealt{vanderWalt2011}),
\texttt{matplotlib} (\citealt{Hunter2007}), 
\texttt{pandas} (\citealt{McKinney2010}),
\revision{
\texttt{casutools} (\citealt{Irwin2004}),
\texttt{banyan sigma} (\citealt{Gagne2018}),
\texttt{SPOC pipeline} (\citealt{Jenkins2016}),
\texttt{stella} (\citealt{Feinstein2020,stella-code}),
\texttt{everest} (\citealt{Luger2016,Luger2018}),
\texttt{processing} (\url{https://processing.org/})
}

\section*{Appendix}
\label{s:Appendix}
\renewcommand\thefigure{A\arabic{figure}}   
\setcounter{figure}{0} 
\renewcommand\thetable{A\arabic{table}}   
\setcounter{table}{0}    


\begin{figure*}[!htbp]
    \centering
    \includegraphics[width=0.9\textwidth]{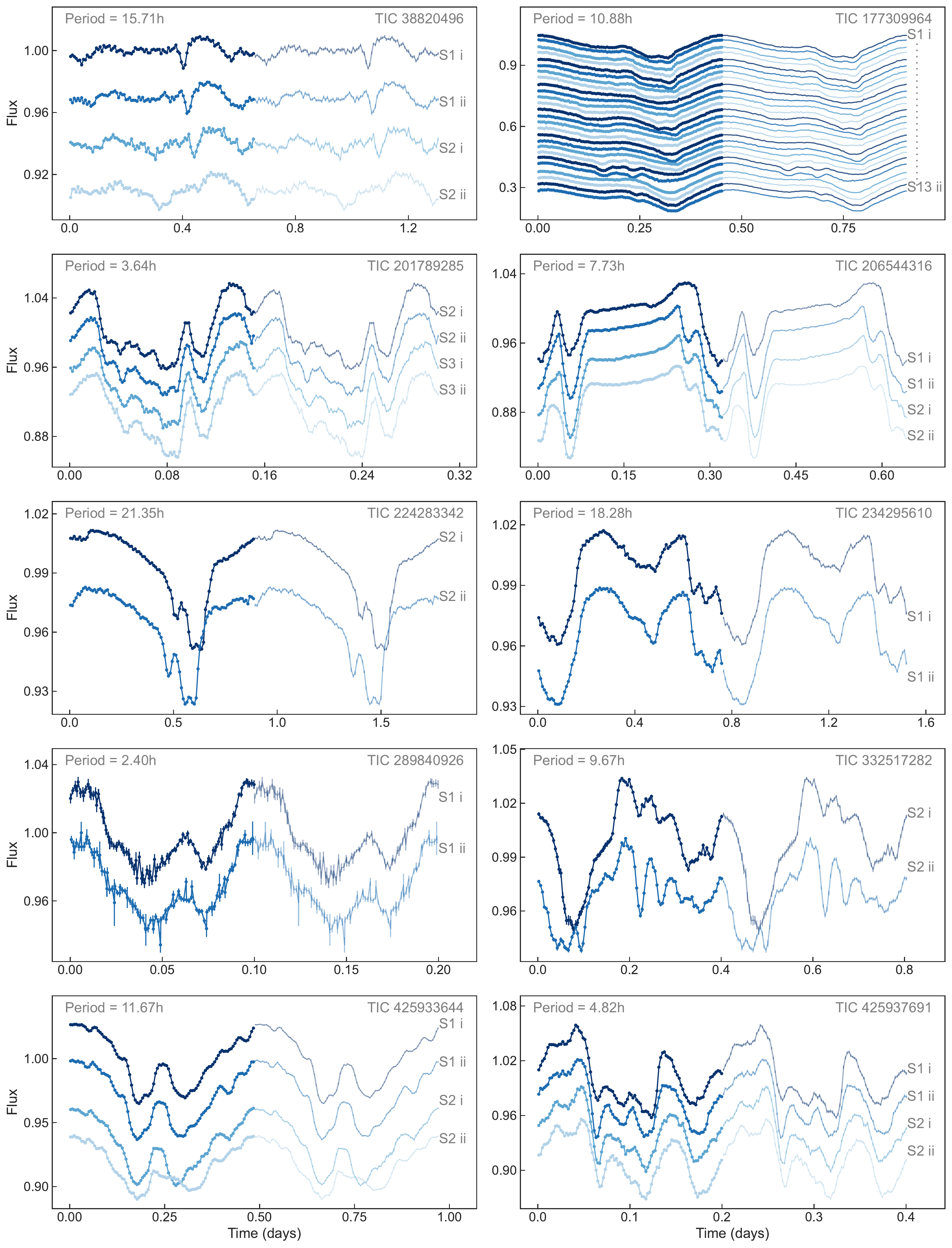}
    \caption{All available TESS light curves of young M dwarfs with complex photometric variability, spanning multiple Sectors of TESS observations. 
    Each light curve shows the data of one TESS orbit (14\,days of observations) phase-folded onto the rotational period of each star, with, e.g., "S1 i" denoting the first orbit of Sector 1 and "S13 ii" denoting the second orbit of Sector 13.
    Flares have been removed prior to phase-folding and plotting. All complex rotators' features show remarkable stability and longevity over many weeks to a year (at least). See also Fig.~\ref{fig:longevity} for a closer view of TIC 17730996.}
    \label{fig:TESS_lightcurves_collage}
\end{figure*}

\begin{figure*}[!htbp]
    \centering
    \includegraphics[width=0.7\textwidth]{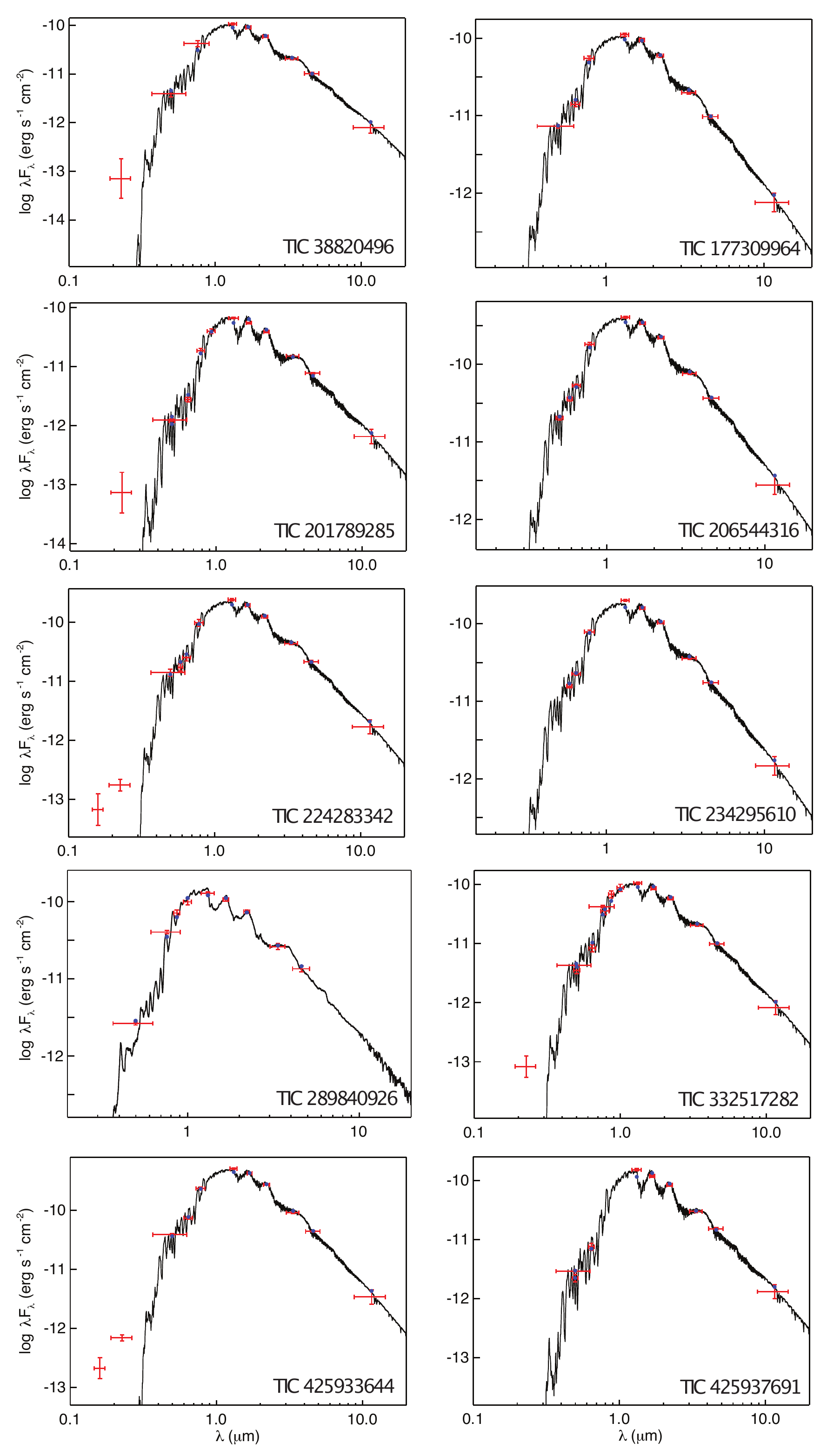}
    \caption{All spectral energy distribution (SED) fits for the ten \textit{complex rotators}. The results are used to determine the stars' apparent effective temperatures and radii. 
    The near-UV and blue optical excesses for some targets are consistent with chromospheric activity, another indicator of strong magnetic activity for these stars.
    Notably, no infrared excesses are seen.}
    \label{fig:SED_collage}
\end{figure*}



\begin{figure*}[!htbp]
    \centering
    \includegraphics[width=\columnwidth]{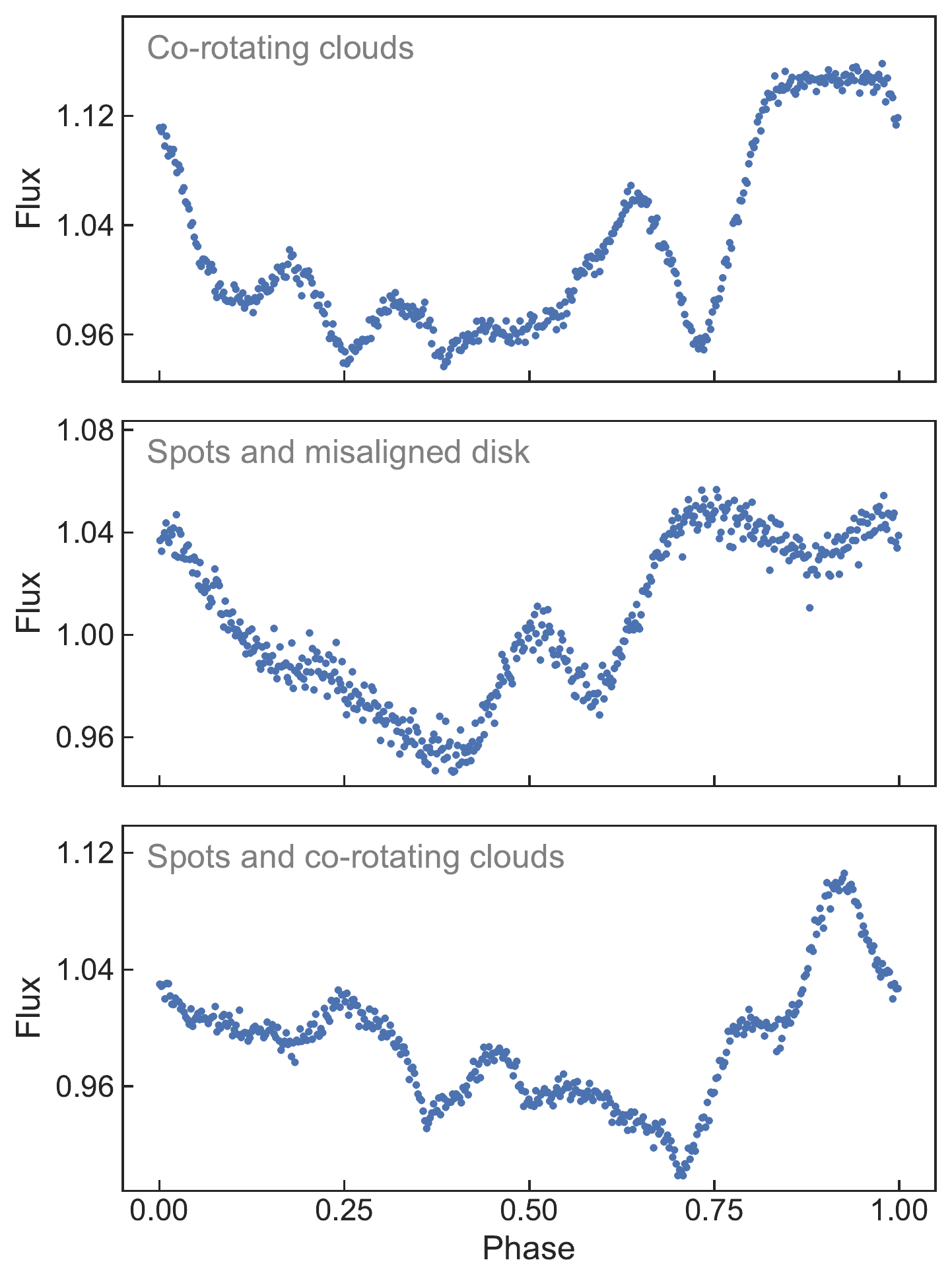}
    \caption{
        \revision{Toy models of a \textit{complex rotator} derived from three different hypotheses, aiming to replicate the morphology of TIC~201789285.}
    }
    \label{fig:toy_models}
\end{figure*}

\newpage
\begin{table*}[!htbp]
    \centering
    \hspace*{-1.6cm}\begin{tabular}{lcccccc}
         \hline
         \hline
         Target & Telescope & Date &  Filter & Exposure (s) & On sky duration (h) & \# images \\
         \hline
         TIC~201789285 & SSO/Io & 20191107 & r' & 35 & 4.93 & 392\\
                       & SSO/Europa & 20191107 & z' & 40 & 4.94 & 357\\
         \hline
         TIC~206544316 & SSO/Io & 20191205 & z' & 10 & 6.23 & 1096\\
               & SSO/Europa & 20191205 & i' & 10 & 6.23 & 1101 \\
               & SSO/Ganymede & 20191205 & r' & 30 & 6.22 & 560 \\
         \hline
         TIC~332517282 & SSO/Io & 20191220 & r' & 60 & 2.60 & 135 \\
              & SSO/Europa & 20191220 & z' & 12 & 2.63 & 431 \\
        \hline
         TIC~425933644  & SSO/Callisto & 20191213 & g' & 50 & 4.62 & 278 \\
              & SSO/Europa & 20191213 & z' & 10 & 4.64 & 827\\
        \hline
    \end{tabular}
    \caption{Summary of all SSO observations for four of the ten \textit{complex rotators}.}
    \label{tab:sso_observations}
\end{table*}


\clearpage
\bibliography{references}


\end{document}